\documentclass[%
 reprint,
 superscriptaddress,
 amsmath,amssymb,
 aps,
]{revtex4-2}

\usepackage{graphicx}
\usepackage{dcolumn}
\usepackage{bm}
\usepackage{epstopdf}
\usepackage{amssymb}
\usepackage{amsmath}
\usepackage{cancel}
\usepackage{color}
\usepackage{siunitx}
\usepackage{nicefrac}
\usepackage{braket}
\usepackage{physics}
\usepackage{mathtools}
\usepackage{epstopdf}
\usepackage{verbatim}
\usepackage{soul}
\usepackage{xr}
\usepackage{xcolor}
\usepackage{float}
\usepackage{multirow}
\usepackage{setspace}

\usepackage{times}    
\usepackage{newtxmath} 

\usepackage[toc,page]{appendix}
\usepackage[section]{placeins}
\usepackage[utf8]{inputenc}
\usepackage[colorlinks]{hyperref}

\DeclareUnicodeCharacter{2009}{\,}

\begin{document}

\preprint{APS/123-QED}

\title{Robust multi-qubit quantum network node with integrated error detection}

\author{P.-J. Stas}
 \thanks{These authors contributed equally to this work.}
 \affiliation{Department of Physics, Harvard University, Cambridge, Massachusetts 02138, USA}
\author{Y. Q. Huan}
 \thanks{These authors contributed equally to this work.}
 \affiliation{Department of Physics, Harvard University, Cambridge, Massachusetts 02138, USA}
\author{B. Machielse}
 \thanks{These authors contributed equally to this work.}
 \affiliation{Department of Physics, Harvard University, Cambridge, Massachusetts 02138, USA}
 \affiliation{AWS Center for Quantum Networking, Boston, Massachusetts 02210, USA}
\author{E. N. Knall}
 \affiliation{John A. Paulson School of Engineering and Applied Sciences, Harvard University, Cambridge, Massachusetts 02138, USA}
\author{A. Suleymanzade}
 \affiliation{Department of Physics, Harvard University, Cambridge, Massachusetts 02138, USA}
\author{B. Pingault}
 \affiliation{John A. Paulson School of Engineering and Applied Sciences, Harvard University, Cambridge, Massachusetts 02138, USA}
 \affiliation{QuTech, Delft University of Technology, 2600 GA Delft, The Netherlands}
 \affiliation{Kavli Institute of Nanoscience Delft, Delft University of Technology,
 2600 GA Delft, The Netherlands}
\author{M. Sutula}
 \affiliation{Department of Physics, Harvard University, Cambridge, Massachusetts 02138, USA}
\author{S. W. Ding}
 \affiliation{John A. Paulson School of Engineering and Applied Sciences, Harvard University, Cambridge, Massachusetts 02138, USA}
\author{C. M. Knaut}
 \affiliation{Department of Physics, Harvard University, Cambridge, Massachusetts 02138, USA}
\author{D. R. Assumpcao}
 \affiliation{John A. Paulson School of Engineering and Applied Sciences, Harvard University, Cambridge, Massachusetts 02138, USA}
\author{Y.-C. Wei}
 \affiliation{Department of Physics, Harvard University, Cambridge, Massachusetts 02138, USA}
\author{M. K. Bhaskar}
 \affiliation{Department of Physics, Harvard University, Cambridge, Massachusetts 02138, USA}
 \affiliation{AWS Center for Quantum Networking, Boston, Massachusetts 02210, USA}
\author{R. Riedinger}
 \affiliation{Department of Physics, Harvard University, Cambridge, Massachusetts 02138, USA}
 \affiliation{Institut f\"ur Laserphysik und Zentrum f\"ur Optische Quantentechnologien,
 Universit\"at Hamburg, 22761 Hamburg, Germany}
 \affiliation{The Hamburg Centre for Ultrafast Imaging, 22761 Hamburg, Germany}
\author{D. D. Sukachev}
 \affiliation{Department of Physics, Harvard University, Cambridge, Massachusetts 02138, USA}
 \affiliation{AWS Center for Quantum Networking, Boston, Massachusetts 02210, USA}
\author{H. Park}
 \affiliation{Department of Physics, Harvard University, Cambridge, Massachusetts 02138, USA}
 \affiliation{Department of Chemistry and Chemical Biology, Harvard University,
 Cambridge, Massachusetts 02138, USA}
\author{M. Lon{\v{c}}ar}
 \affiliation{John A. Paulson School of Engineering and Applied Sciences, Harvard University, Cambridge, Massachusetts 02138, USA}
\author{D. S. Levonian}
 \affiliation{Department of Physics, Harvard University, Cambridge, Massachusetts 02138, USA}
 \affiliation{AWS Center for Quantum Networking, Boston, Massachusetts 02210, USA}
\author{M. D. Lukin}
 \altaffiliation{Corresponding author. E-mail:  lukin@physics.harvard.edu.}
 \affiliation{Department of Physics, Harvard University, Cambridge, Massachusetts 02138, USA}
 
\date{\today}

\begin{abstract}
Long-distance quantum communication and networking require quantum memory nodes with efficient optical interfaces and long memory times. We report the realization of an integrated two-qubit network node based on silicon-vacancy centers (SiVs) in diamond nanophotonic cavities. Our qubit register consists of the SiV electron spin acting as a communication qubit and the strongly coupled $^{29}$Si nuclear spin acting as a memory qubit with a quantum memory time exceeding two seconds. By using a highly strained SiV with suppressed electron spin-phonon interactions, we realize electron-photon entangling gates at elevated temperatures up to 1.5 K and nucleus-photon entangling gates up to 4.3 K. Finally, we demonstrate efficient error detection in nuclear spin-photon gates by using the electron spin as a flag qubit, making this platform a promising candidate for scalable quantum repeaters.
\end{abstract}

\maketitle

The ability to distribute  quantum information over extended distances  \cite{Kimble2008,Wehner2018} constitutes an important enabling technology in quantum information science, with applications in quantum key distribution \cite{Ekert1991,Lo2014}, nonlocal sensing \cite{Khabiboulline2019}, and distributed quantum computation~\cite{Buhrman2003,Cuomo2020}. A key requirement for the realization of such long-distance quantum networking involves the development of quantum repeaters \cite{Dr1999, Jiang2009} to mitigate photonic qubit loss during the transmission over extended distances. Proposed repeater architectures require nodes containing multiple qubits that can collect, store, and process information communicated via photonic channels \cite{Childress2006_PRL}.

Color centers in diamond nanophotonic structures \cite{Nguyen2019_PRB,Schmidgall2018,Rugar2020,Guo2021} have recently emerged as promising candidates for realizing such nodes due to their long coherence times, high-fidelity single-qubit gates, and efficient photonic interfaces. The integration of all these features into a single device has led to the demonstration of memory-enhanced quantum communication \cite{Bhaskar2020}. However, scalable implementation of quantum repeaters using this approach requires deterministic access to additional long-lived quantum memory qubits to perform entanglement swapping, purification and error detection \cite{Dr1999, Childress2006_PRL}.  Furthermore, the technical complexity and costs associated with the need for operation at dilution refrigerator temperatures to avoid phonon-induced dephasing \cite{Jahnke2015,Becker2018} has stimulated efforts to identify quantum memory systems capable of operating within a wider range of environmental conditions \cite{Rose2018,Trusheim2020}.

In this Report, we address these challenges by using a highly strained silicon-vacancy (SiV) color center~\cite{Jahnke2015,Meesala2018} featuring the $^{29}$Si isotope, with its nuclear spin \cite{Pingault2017} serving as a deterministic long-lived memory qubit. We demonstrate full control of this two-qubit register with a $^{29}$Si quantum memory storage time of over 2 seconds. Selective qubit readout via a novel phase-based readout method enables multiple electron state resets before deterioration of the nuclear spin memory. Using this system, we achieve electron spin-photon entanglement at temperatures up to 1.5 K and direct nuclear-photon entanglement up to 4 K. Finally, we combine these elements to demonstrate nuclear-photon spin entanglement with integrated error detection using the electronic spin as a flag qubit to detect operational errors. 

\begin{figure}
\includegraphics[trim=0cm 15.5cm 10cm 0cm, width=\linewidth]{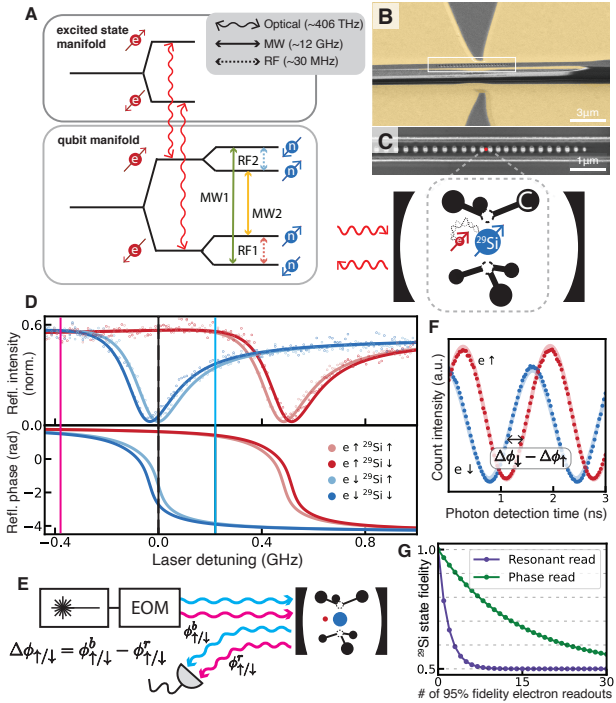}
\caption{\textbf{Quantum network node based on $^{29}$SiV.} (\textbf{A}) Energy level diagram of the $^{29}$SiV. Allowed magnetic dipole transitions are shown by arrows in the qubit manifold for the case when the $\mathbf{B}$-field is oriented along the symmetry axis. (\textbf{B}) False color scanning electron microscope (SEM) image of the device, showing the gold coplanar waveguides in yellow. (\textbf{C}) Zoomed in SEM image of the nanophotonic cavity in which the $^{29}$SiV is located. (\textbf{D}) Spin-dependent reflection intensity (top) and phase (bottom) as a function of laser frequency. (\textbf{E}) Experimental setup for phase readout of the electron spin state. (\textbf{F}) Measured spin-dependent phase shift of the beating pattern between sidebands. (\textbf{G}) Coherence of a nuclear superposition state as a function of the number of electron readouts for resonant and phase readouts.}
\label{fig:structure}
\end{figure}

The $^{29}$SiV in an external magnetic field constitutes a two-qubit system of four spin states with nondegenerate transition frequencies (Fig. \ref{fig:structure}A). The electron and nuclear spin qubits are coherently controlled using microwave (MW) and radiofrequency (RF) pulses, respectively, delivered via gold coplanar waveguides (Fig. \ref{fig:structure}B). The SiV is embedded in a nanophotonic cavity (Fig. \ref{fig:structure}C) enhancing optical transitions to the excited state manifold at 737 nm (Fig. \ref{fig:structure}A, wavy arrows) used for state readout and spin-photon entanglement. The SiV-cavity system exhibits high-contrast spin-dependent reflection spectra (Fig. \ref{fig:structure}D) enabled by the strong cavity coupling (cooperativity $C = 1.6$ \cite{SI}). 

High-fidelity resonant readout of the electron state is achieved by measuring the reflected intensity of a laser at the frequency of maximum reflection intensity contrast (black line in Fig. \ref{fig:structure}D) \cite{Nguyen2019_PRB,Levonian2022}. However, the nuclear qubit experiences dephasing from the laser during readout, with the highest decoherence rate occurring when the laser is near the resonant readout frequency \cite{SI}. To enable selective readout of the electronic spin qubit, we probe the system at laser frequencies where minimal nuclear dephasing occurs and make use of the electron spin-dependent phase -- instead of intensity -- of reflected photons for readout. Using an electro-optic modulator (EOM) to generate sidebands (Fig. \ref{fig:structure}E), we send two tones (blue and magenta lines in Fig. \ref{fig:structure}D) along the same path and measure the spin-dependent reflected phase difference via the resulting beating pattern to determine the electron state (Fig. \ref{fig:structure}F) \cite{SI}. With this phase-based readout, the electron can be read 14 times at 95\% fidelity before causing a $1/e$ loss of nuclear coherence. This constitutes an 8-fold improvement over the resonant readout method (Fig. \ref{fig:structure}G), with further improvements expected through cavity design and magnetic field optimization \cite{SI}. 

\begin{figure}
\includegraphics[trim=0cm 18.5cm 11.8cm 0cm, width=\linewidth]{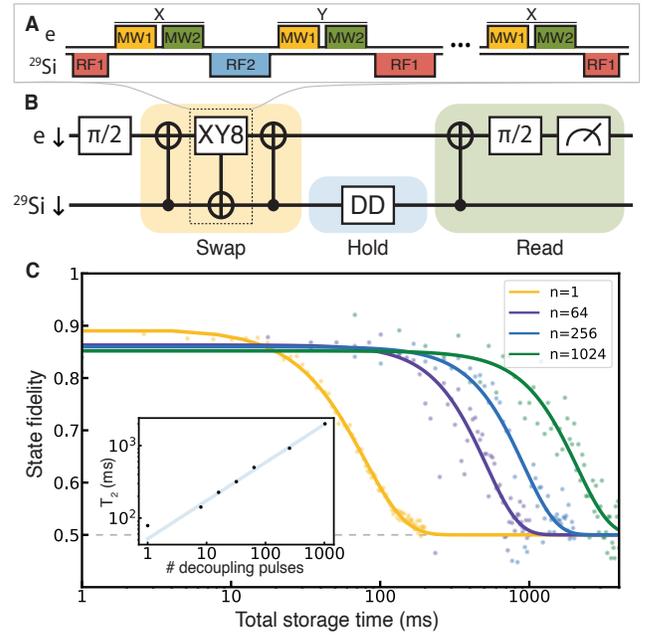}
\caption{\textbf{Long-lived quantum memory based on $^{29}$Si nuclear spin.} (\textbf{A}) Pulse sequence for a decoupled C$_e$NOT$_n$ gate. (\textbf{B}) Sequence for swapping the electron state onto the nucleus, after which a nuclear XY8 decoupling sequence with $n=1, 64, 256, 1024$ pulses is applied and the nuclear state is measured through resonant readout of the electron. (\textbf{C}) Measured nuclear state fidelity after sequence (B). Inset: $T_{2,n}$ as a function of the number of decoupling pulses $n$ fitted to $T_{2,n} \propto n^\alpha$ where $\alpha=0.53$.} 
\label{fig:dcpld-CNOT}
\end{figure}

The two-qubit $^{29}$SiV system is fully controlled by selectively driving the four single-spin-flipping transitions (Fig. \ref{fig:structure}A) to implement the four possible controlled-NOT (CNOT) gates: two electron-flipping gates C$_n$NOT$_e$ (MW1) and $\overline{\mathrm{C}_n\mathrm{NOT}_e}$ (MW2), and two nucleus-flipping gates C$_e$NOT$_n$ (RF1) and $\overline{\mathrm{C}_e\mathrm{NOT}_n}$ (RF2), where the absence (presence) of the overbar indicates conditioning on the control spin being in the down (up) state. We measure a fidelity of $99.9\pm0.1\%$ for the C$_n$NOT$_e$ gate with a gate time of 30.0 ns. As the time to drive a nuclear $\pi$ rotation (\SI{23}{\micro\second}) is longer than the electron dephasing time $T_{2,e}^*$ (\SI{5}{\micro\second}) \cite{SI}, transfer of electron superposition states onto the nucleus is not possible with direct driving for the C$_e$NOT$_n$ gate. To circumvent this problem, we increase the electron coherence time using a dynamical decoupling sequence and interleave it with step-wise nuclear rotations~\cite{Bradley2019,vanderSar2012} to implement a decoupled C$_e$NOT$_n$ (Fig. \ref{fig:dcpld-CNOT}A). Specifically, we apply RF1 in the first window, and alternate it with applying RF2 after every subsequent unconditional electron $\pi$ pulse to account for the flipping electron state while keeping the nuclear rotation conditional on the same initial electron state, achieving a gate fidelity of $93.7\pm0.7\%$ \cite{SI} within a gate time of \SI{29.2}{\micro\second}. With the decoupled C$_e$NOT$_n$ we can swap a superposition state from the electron onto the nuclear spin state (Fig. \ref{fig:dcpld-CNOT}B), where it can be stored using an XY8 decoupling sequence. We find that the nuclear coherence time scales with $n$, the number of decoupling pulses applied, as $T_{2,n} \propto n^{0.53}$ with a maximum measured nuclear memory time of $T_{2,n} = 2.1\pm0.1$ seconds (Fig. \ref{fig:dcpld-CNOT}C) when using a 128-XY8 sequence.

\begin{figure*}[t]
\includegraphics[trim=0cm 17cm 8cm 0cm]{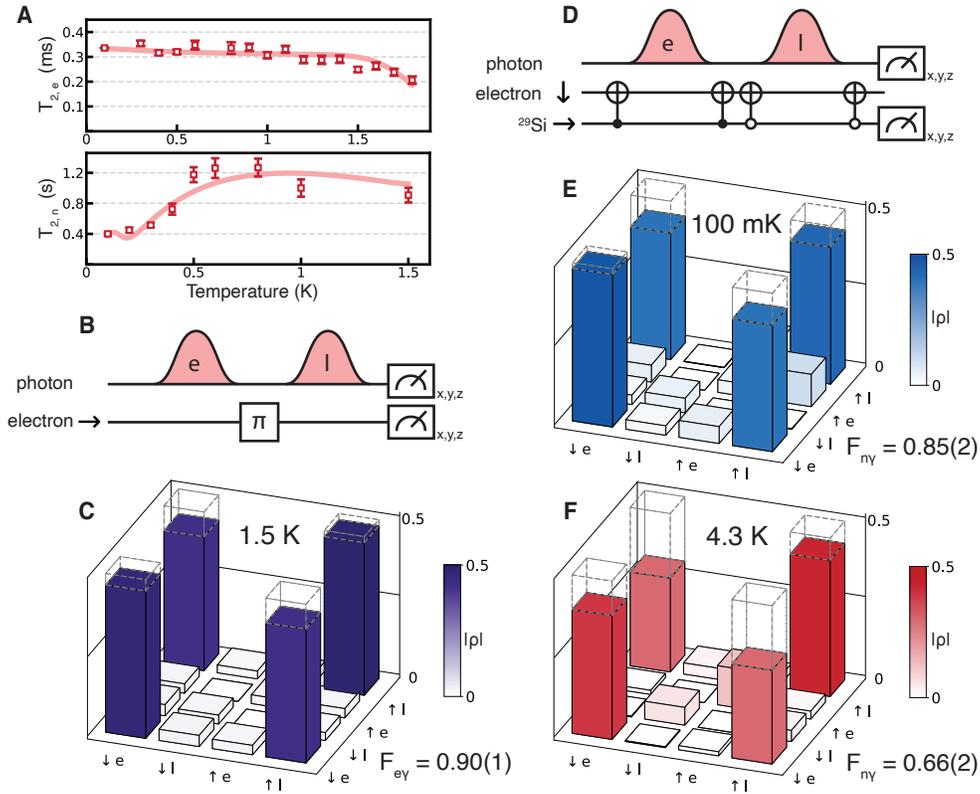}
\caption{\textbf{Spin-photon entanglement at elevated temperatures.} (\textbf{A}) Coherence time of the electron (top) and nucleus (bottom) obtained using 8-XY8 sequences as a function of temperature. Initial increase of the nuclear $T_{2,n}$ is due to motional averaging of the noise bath. Solid lines are fits to decoherence models of the spin environment described in \cite{SI}. (\textbf{B}) Electron spin-photon gate implementation with (\textbf{C}) reconstructed density matrix at 1.5 K using resonant electron readout. (\textbf{D}) Circuit diagram for the PHOton-Nucleus Entangling (PHONE) gate. CNOT gates with black and white control dots refer to $\mathrm{C}_n\mathrm{NOT}_e$ and $\overline{\mathrm{C}_n\mathrm{NOT}_e}$ gates respectively. Reconstructed density matrices for photon-nuclear spin entanglement at (\textbf{E}) 100 mK and (\textbf{F}) 4.3 K.}
\label{fig:spin-photon-gates}
\end{figure*}

To enable robust spin-photon entanglement, we use a SiV with large residual strain (ground state splitting $\Delta_{GS} = 554$ GHz \cite{SI}), which greatly suppresses the rate of thermal decoherence processes \cite{Meesala2018}, enabling  operation at 1.5 K without a significant reduction in $T_{2,e}$ (Fig. \ref{fig:spin-photon-gates}A). We implement high-temperature spin-photon entangling gates between the electron spin and time-bin qubits in the $\lbrace \ket{e}, \ket{l} \rbrace$ basis, corresponding to the presence of a photon in either the early or late time-bin. The entangling gate sequence \cite{Nguyen2019_PRB,Bhaskar2020} shown in Fig. \ref{fig:spin-photon-gates}B generates the entangled state $(\ket{e\downarrow_e} + \ket{l \uparrow_e})/\sqrt{2}$ conditioned on the detection of a single reflected photon. We find a Bell state fidelity of $F_{e\gamma} = 0.91 \pm 0.02$ ($0.90 \pm 0.01$) at 0.1 K (1.5 K) (Fig. \ref{fig:spin-photon-gates}C), which is primarily limited by residual reflections from the $\ket{\downarrow_e}$ state and imperfect photonic state measurements.

To further extend the entanglement capabilities of our spin-photon interface, we introduce a novel PHOton-Nucleus Entangling (PHONE) gate that directly entangles the $^{29}$Si nuclear spin with a photonic qubit using only fast MW gates and allows operating temperatures up to 4.3 K. In this scheme, the photon-nucleus-electron system is initialized in the $(\ket{e} + \ket{l}) (\ket{\uparrow_n} + \ket{\downarrow_n}) \ket{\downarrow_e} /2$ state and we apply a sequence of CNOTs as shown in Fig. \ref{fig:spin-photon-gates}D such that the SiV reflectivity in each time bin is conditional on the nuclear state; the final entangled state is given by $(\ket{e \uparrow_n} + \ket{l \downarrow_n})\ket{\downarrow_e}/\sqrt{2}$. We measure the resulting Bell state fidelity to be $F_{n\gamma} = 0.85 \pm 0.02$ ($0.66 \pm 0.02$) at 0.1 K (4.3 K) (Fig. \ref{fig:spin-photon-gates}E, F), with dominant infidelities due to MW gate errors and the nuclear state-dependence of the optical transition frequency which prevents the photon frequency from being tuned to maximize contrast for both nuclear spin states simultaneously (Fig. \ref{fig:structure}D) \cite{SI}.

\begin{figure}
\includegraphics[trim=0cm 20cm 13.6cm 0cm,width=\linewidth]{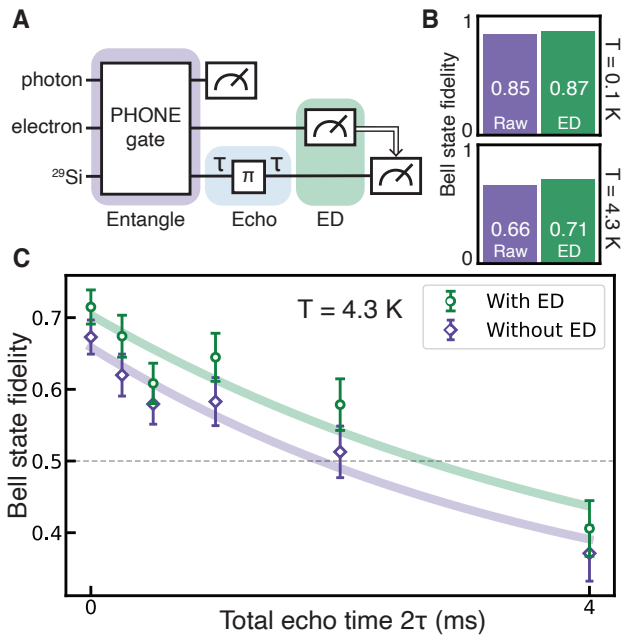}
\caption{\textbf{Spin-photon entanglement with integrated error detection.} (\textbf{A}) Gate sequence for entanglement between a photon and nucleus and subsequent state storage with error detection (ED) based on resonant readout of the electronic spin. (\textbf{B}) Bell state fidelity improvement with ED at 0.1 K and 4.3 K. (\textbf{C}) Bell state fidelity as a function of storage time for the sequence shown in (A) before and after error detection. Fits are exponential curves decaying to the maximally mixed state fidelity $\bra{\Phi^+}\rho_{\mathrm{mm}}\ket{\Phi^+} = 0.25$ as the nuclear lifetime $T_{1,n}$ is comparable to the coherence time; fitted decay constants are 4.5 (3.8) ms with (without) ED.} 
\label{fig:error-detection}
\end{figure}

A key feature of the PHONE gate is that the electron should always remain in the $\ket{\downarrow_e}$ state after a successful gate application. Since the electron spin mediates the interface between the photon and the nucleus, electron spin flips can be used as an integrated error witness to detect gate errors. Similar to flag qubits in error correction protocols \cite{Chao2018}, PHONE gate errors can be reduced at the cost of some gate failure probability \cite{Borregaard2015} by measuring the state of the electron spin qubit (Fig. \ref{fig:error-detection}A). Specifically, if the electron spin is measured to be in the $\ket{\uparrow_e}$ state following a PHONE gate application, the prepared photon-nuclear spin entangled state is discarded. As shown in Fig. \ref{fig:error-detection}B, the use of the procedure results in a Bell state fidelity increase of 2\% (7\%) to $F_{n\gamma} = 0.87 \pm 0.02$ ($0.71 \pm 0.02$) with an error detection rate of 8.4\% (13.9\%) at 0.1 K (4.3 K). In Fig. \ref{fig:error-detection}C we combine all these components to implement a PHONE gate at 4.3 K with error detection and store the spin-photon entangled state in the nuclear spin memory using an echo sequence (Fig. \ref{fig:error-detection}A) for over 2.5 ms above the threshold value of 50\%, as compared to 1.5 ms without error detection. The error detection protocol detects MW gate errors and $T_{1,e}$ and $T_{2,e}$-limited depolarization and dephasing, and the large gain in fidelity at 4.3~K is due to the greater contribution of the detectable errors from short $T_{1,e}$ and $T_{2,e}$ times at this temperature \cite{SI}. Conversely, the improvement from error detection at 0.1~K is limited as the infidelity is dominated by non-detectable errors from photon state measurement and SiV optical contrast.  

These observations open up several new avenues for realizing quantum networks and exploring their applications.
The access to an additional memory qubit directly enables improved memory-enhanced quantum key distribution \cite{Bhaskar2020} by extending the time window for photons arriving from different communicating parties to the memory node for asynchronous Bell measurement, whereas the phase readout protocol facilitates electron resets between entanglement attempts while information is stored on the nucleus. A nuclear memory can additionally expand the capabilities of the SiV as a single photon source for the creation of photonic cluster states \cite{Lindner2009,Knall2022}.
The methods demonstrated here can also enable the deployment of scalable SiV-based quantum repeater networks. Based on a large-scale survey of a second chip fabricated using the current process \cite{SI}, we find that more than 11\% of all SiVs are sufficiently strained for operation at 1.5 K, indicating the possibility for a significant number of highly-strained devices in nanophotonic cavities capable of operating at elevated temperatures. Moreover, recently demonstrated techniques for strain tuning of nanocavities \cite{Machielse2019} can enable fully deterministic access to high-strain operation, eliminating the need for dilution refrigerators. 
The fidelity gain from our error detection can be more significant for more complex spin-photon entangling sequences, such as PHONE-type gates entangling successive photons with the nucleus or entangling a photon with multiple nuclei strongly coupled to the SiV. It also opens opportunities to operate more effectively in regimes where the SiV electron coherence properties deteriorate, including at higher temperatures as demonstrated here, and in a misaligned magnetic field as is required for acoustic spin control \cite{Maity2020} and single photon generation \cite{Knall2022}. The use of cavities with higher cooperativity as demonstrated previously \cite{Bhaskar2020} should allow for higher fidelity and efficiency of electron-spin photon and PHONE gates, as well as improved $^{29}$Si state preservation during electron readout~\cite{SI}. Finally, nearby $^{13}$C spins, such as one associated with the SiV used in this work \cite{SI}, can be used as additional memory resources \cite{Bradley2019,Kalb2017}. Apart from realizing multi-node quantum network protocols \cite{Dr1999, Jiang2009, Childress2006_PRL}, these systems can also allow for the generation of complex photonic tree cluster states enabling robust one-way long-distance quantum communication \cite{Borregaard2020}.

\bibliography{main.bib}

\textbf{Acknowledgments:} We thank J. Borregaard for discussions, and J. MacArthur for assistance with electronics. This work was supported by the NSF, CUA, DoD/ARO DURIP, AFOSR MURI, and DOE (award no. DE-SC0020115). Devices were fabricated at Harvard CNS, NSF award no. 1541959. Y.Q.H acknowledges support from the A*STAR National Science Scholarship. D.A., E.N.K., and B.M. acknowledge support from an NSF GRFP No. DGE1745303. B.P. acknowledges support through a Marie Sk\l{}odowska-Curie fellowship from the European Union's Horizon 2020 research and innovation programme under the Grant Agreement No. 840968 (COHESiV). M.S. acknowledges funding from the NASA Space Technology Graduate Research Fellowship Program. R. R. acknowledges support from the Alexander von Humboldt Foundation and the Cluster of Excellence `Advanced Imaging of Matter' of the Deutsche Forschungsgemeinschaft (DFG) - EXC 2056 - project ID 390715994.

\onecolumngrid


\setcounter{equation}{0}
\setcounter{figure}{0}
\setcounter{table}{0}
\setcounter{page}{1}
\setcounter{section}{0}
\makeatletter
\renewcommand{\theequation}{S\arabic{equation}}
\renewcommand{\thefigure}{S\arabic{figure}}
\renewcommand{\thetable}{S\arabic{table}}

\pagebreak
\begin{center}
\Large Supplemental Materials for\\\textbf{Robust multi-qubit quantum network node with integrated error detection}
\end{center}

\singlespacing

\section{Sample Fabrication}

The nanophotonic crystal cavities utilized in this manuscript were patterned out of bulk diamond utilizing techniques previously reported in \cite{Nguyen2019_PRB}. Subsequent implantation of $^{29}$Si through apertures patterned into a resist mask, followed by high temperature annealing, enables incorporation of color centers into these photonic crystals with a high degree of spatial accuracy. The optical properties of these devices can be measured through a hydrofluoric acid wet-etched tapered optical fiber \cite{Burek2017} brought into contact with the tapered end of the diamond devices. By measuring the reflection spectra of the photonic crystal devices inside a dilution refrigerator operating at 100 mK, strong modulation of the cavity properties by the presence of these color centers can be observed, indicating relatively high cooperativity ($C=4g^{2}/\kappa\gamma= 1.6$; see section on cavity-QED model for parameter values) coupling of the emitters with photons in the cavity mode. Further improvements of cooperativities can be achieved through optimization of the etch mask \cite{Chia22}. Candidate devices can then be identified and, after removal from the cryostat, deposited with coplanar waveguides which are patterned using electron beam lithography (Fig. \ref{fig:structure}B). 

\section{Experimental setup}

We perform all measurements in a dilution refrigerator (BlueFors BF-LD250) with an effective base temperature of 100 mK. The dilution refrigerator is equipped with a superconducting vector magnet (American Magnets Inc. 6-1-1 T), a home-built free-space wide-field microscope with an objective lens, piezo positioners (Attocube ANPx101 and ANPx311 series), and fiber and microwave feedthroughs (gold coplanar waveguides). A gas line with attached heaters is used for tuning of the nanocavity resonance through nitrogen gas condensation. 

The optical path consists of two separate 737 nm Ti:Sapphire lasers
(TiSaph, M Squared SolsTiS). One of these lasers is used exclusively for phase readout and is tuned to the phase readout carrier frequency, while the other is tuned to the resonant readout frequency and is used for resonant readout, photonic qubit generation, and time delay interferometer (TDI) locking. A separate green diode laser (532 nm, Thorlabs) is used to reset the SiV electron charge state when it ionizes. The acousto-optic modulators (AOMs) on the optical path are Gooch \& Housego R1520 and the EOMs are EOSpace AZ-OK5-10-PFA-PFA-637 Mach-Zehnder amplitude modulators. All of the laser paths (with the exception of the TDI locking beam) are combined into a single-mode S630-HP fiber (Thorlabs) and sent into the dilution fridge where it couples light adiabatically in and out of the diamond nanocavity via a tapered fiber (Thorlabs 630-HP). Reflected photons from the device travel along the same path through a 99:1 beam splitter where it is collected either on an SNSPD (Photon Spot) or a pair of avalanche photodiodes (APDs, Excelitas SPCM-780-14-FC). The APDs are used for all experiments at 4 K and the spin-photon entangling gates; the SNSPDs are used for all other experiments. Counts from these photon detectors are recorded on a time-tagger (Swabian Instruments Time Tagger Ultra) as well as an integrated counter on the arbitrary waveform generator (AWG) for state readout and conditional logic. A pair of photodiodes are placed along the optical path to calibrate the coupling efficiency of the fiber with the device by measuring the ratio of incoming to reflected light.

\begin{figure}[ht]
\centering
\includegraphics[trim=0cm 21cm 5cm 0cm]{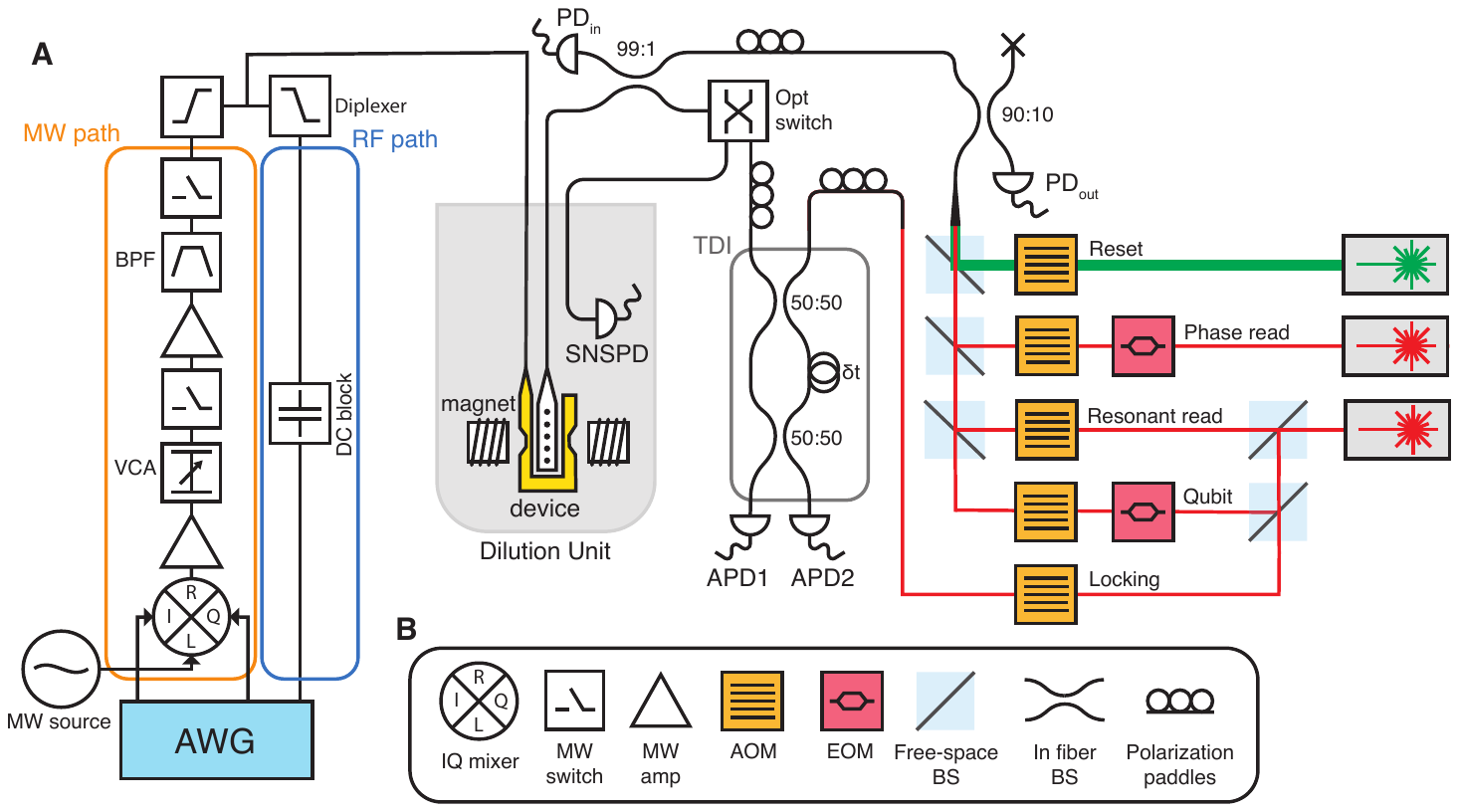}
\caption{(\textbf{A}) Diagram of the setup for optical and microwave control of our system.  (\textbf{B}) Legend of components in (A).}
\label{SIfig:setup}
\end{figure}

The MW and RF chain is driven using a Zurich Instruments HDAWG8 2.4 GSa/s AWG that is also used to time the experimental sequence, switch the AOMs, and generate the qubit gate waveforms on the EOM.  On the microwave path (from bottom to top in Fig. \ref{SIfig:setup}) are an IQ mixer (Marki MMIQ-0218L), a pre-amplifier (ZX60-06183LN+), a voltage controlled attenuator (RFVA T0218A50), a switch (CMCS0947A-C2), an amplifier (ZVE-3W-183+), a bandpass filter (ZVBP-11R7G-S+, $10.70-12.75$ GHz), a high power switch (Qorvo QPC2040), and a diplexer (ZDSS-3G4G-S+). The microwave source used as the LO of the IQ mixer is an Agilent 83732B Synthesized Signal Generator and is set to 11.85 GHz at 14 dBm. The RF pulses are directly synthesized from the HDAWG and sent to the diplexer via a DC block (MCL15542 BLK-18-S+). The microwave and RF signals are sent into the dilution refrigerator via a coaxial cable that is wirebonded to the coplanar waveguides on the diamond chip. All experimental sequences are run using \texttt{pylabnet}, a custom Python data-taking software framework \cite{pylabnet2021}, together with fast logic on the AWG.

\section{Cavity-QED Model}

We use the following expressions for the reflection, transmission, and scattering amplitudes of light for a two-level system coupled to a cavity \cite{Reiserer2015}:

$$ r(\omega) = 1 - \frac{\kappa_{\mathrm{in}}}{i(\omega - \omega_c) + \kappa_{\mathrm{tot}}/2 + g^2 / (i(\omega - \omega_a) + \gamma/2) } $$
$$ t(\omega) = \frac{\sqrt{\kappa_{\mathrm{in}}\kappa_{\mathrm{out}}}}{i(\omega - \omega_c) + \kappa_{\mathrm{tot}}/2 + g^2 / (i(\omega - \omega_a) + \gamma/2) } $$
$$ s(\omega) = \frac{\sqrt{\kappa_{\mathrm{in}}\gamma}g}{(i(\omega - \omega_c) + \kappa_{\mathrm{tot}}/2)(i(\omega - \omega_a) + \gamma/2) + g^2} $$

where $r(\omega)$, $t(\omega)$, and $s(\omega)$ are the reflection, transmission, and scattering amplitudes respectively. The transmitted component is defined to include both light that that is coupled through the transmission port as well as light that is scattered off the cavity due to fabrication imperfections; both modes of losses are incorporated into $\kappa_{\mathrm{out}}$. $\kappa_{\mathrm{in}}$ is the coupling rate of the cavity into the in-coupling port while $\kappa_{\mathrm{tot}} = \kappa_{\mathrm{in}} + \kappa_{\mathrm{out}}$ is the total cavity decay rate. The scattered component is defined as the component that decays into free space due to spontaneous emission, and is characterized by the bare SiV emitter linewidth $\gamma$. Finally, $\omega_{c(a)}$ is the resonance frequency of the cavity (SiV) and $g$ is the coupling rate of the SiV to the cavity. The cooperativity of the system is $C=\frac{4g^2}{\kappa_{tot}\gamma} = 1.6\pm0.1$. Fig. \ref{fig:structure}D in the main text shows the intensity and phase of the reflected light, where the phase is given by $\mathrm{arg}(r(\omega))$ and the intensity by $R(\omega) = | r(\omega)|^2$. The datapoints were fitted to obtain the parameters $\omega_a$ for the 4 transitions together with $\kappa_{\mathrm{in}}$ and $g$. The cavity frequency $\omega_c$ and linewidth $\kappa_{\mathrm{tot}}$ were separately fitted from the bare cavity reflection spectrum.

\begin{table}
\centering
\begin{tabular}{c c c} 
    \hline
    Parameter & Description & Fitted Value \\ 
    \hline\hline
    $\omega_c$ & Cavity resonance frequency & 406.610(1) THz \\
    \hline
    $\omega_{a,\uparrow_e\uparrow_n}$ & Optical resonance frequency for $\ket{\uparrow_e\uparrow_n}$ & 406.678504(2) THz \\
    \hline
    $\omega_{a,\uparrow_e\downarrow}$ & Optical resonance frequency for $\ket{\uparrow_e\downarrow_n}$ & 406.678537(2) THz \\ 
    \hline
    $\omega_{a,\downarrow_e\uparrow_n}$ & Optical resonance frequency for $\ket{\downarrow_e\uparrow_n}$ & 406.679053(2) THz \\
    \hline
    $\omega_{a,\downarrow_e\downarrow_n}$ & Optical resonance frequency for $\ket{\downarrow_e\downarrow_n}$ & 406.679022(2) THz \\
    \hline
    $\kappa_{\mathrm{in}} / 2\pi $ & Coupling rate into cavity & 202(12) GHz \\
    \hline
    $\kappa_{\mathrm{tot}} / 2\pi $ & Total cavity linewidth & 250(14) GHz \\
    \hline
    $g / 2\pi $ & SiV-cavity coupling rate & 3.19(5) GHz \\ 
    \hline
    $\gamma / 2\pi $ & Bare SiV linewidth & 100 MHz \\
    \hline
    \end{tabular}
    \caption{Numerical values of cavity and emitter parameters. The number in parentheses indicate the uncertainty in the last digits. All linewidths and rates are defined as the full-width at half-maximum values. }
    \label{tab:fittedCQED}
\end{table}
    
\section{MW and RF gates}

\begin{table}
\centering
\begin{tabular}{c c} 
    \hline
    Parameter & Value \\ 
    \hline\hline
    $\omega_{MW1}$ & 12.00746 GHz \\
    \hline
    $\omega_{MW2}$ & 12.07384 GHz \\
    \hline
    $\omega_{RF1}$ & 29.636 MHz \\
    \hline
    $\omega_{RF2}$ & 36.614 MHz \\
    \hline
    $\Omega_e / 2\pi$ & 16.7 MHz \\
    \hline
    $\tau_e$ & 30.0 ns \\
    \hline
    $\Omega_n / 2\pi$ & 19.6 kHz (RF1), 24.2 kHz (RF2) \\
    \hline
    $\tau_n$ & \SI{25.5}{\micro\second} (RF1), \SI{20.7}{\micro\second} (RF2) \\
    \hline
    $T_{1,e}$ & 2.9 s (0.1 K), 17 ms (4.3 K) \\
    \hline
    $T_{2,e}^*$ & $\sim$ \SI{5}{\micro\second} (0.1 K)\\
    \hline
    $T_{2,n}^*$ & $\sim$ 5 ms (0.1 K)\\
    \hline
    $T_{2,e}$ with Hahn echo & \SI{78}{\micro\second} (0.1 K), 400 ns (4.3 K) \\
    \hline
    $T_{2,n}$ with Hahn echo & 79 ms (0.1 K), 4.5 ms (4.3 K) \\
    \hline
    Magnetic field & 0.394 T \\
    \hline
    \end{tabular}
    \caption{Microwave and RF driving parameters.}
    \label{tab:pulseParams}
\end{table}

\begin{figure}[ht]
\centering
\includegraphics[trim=0cm 20.5cm 8cm 0cm]{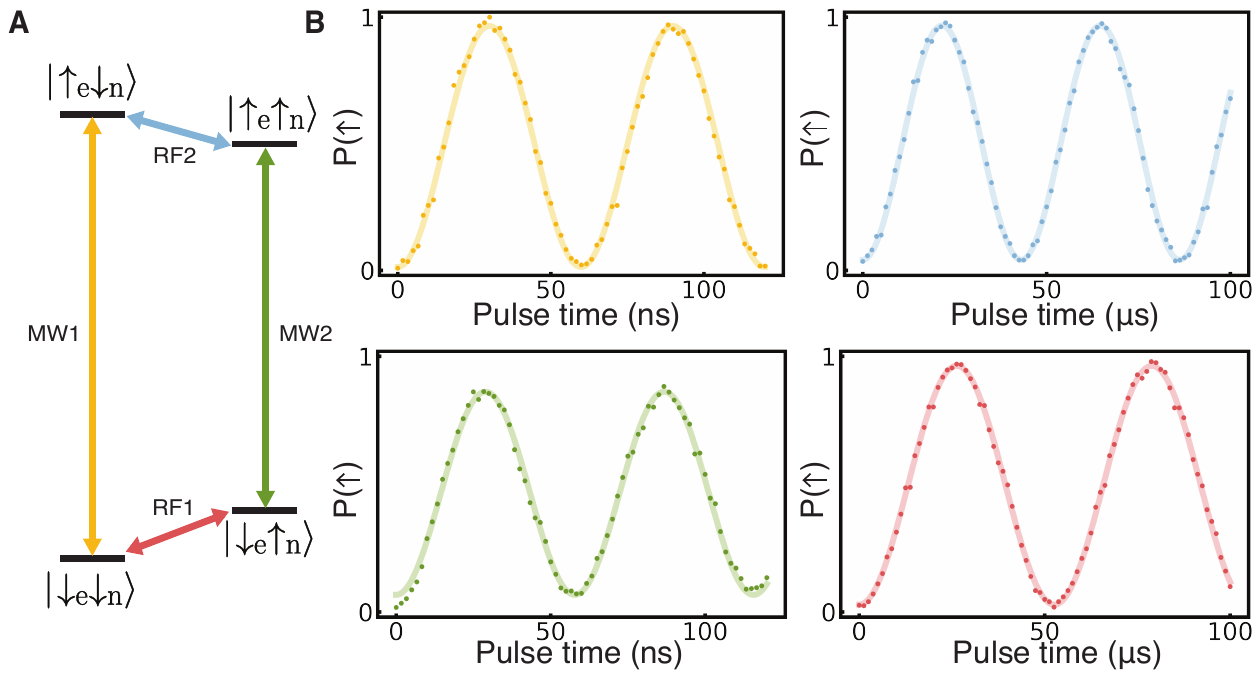}
\caption{(\textbf{A}) Qubit manifold energy level diagram of $^{29}$SiV. Driving on each labeled transition induces a conditional rotation.  (\textbf{B}) Rabi oscillations for each of the four transitions.}
\label{SIfig:rabis}
\end{figure}

All relevant MW and RF transition frequencies as well as pulse parameters are listed in Table \ref{tab:pulseParams}, and Rabi oscillations for the different transitions are shown in Fig. \ref{SIfig:rabis}. The relatively large nuclear Rabi frequency is due to the strong electron-nuclear hyperfine interaction strength. The Rabi frequency $\Omega_e$ and $\pi$-pulse time $\tau_e$ for electron-flipping pulses need to satisfy $\Omega_e \tau_e = \pi$. However, in order to achieve a C$_n$NOT$_e$, we need to ensure that the off-resonant electron transition (i.e. when the nucleus is not in the state being conditioned on) is not flipped by the off-resonant driving. The off-resonant transition is detuned by the $^{29}$Si-electron hyperfine coupling strength $A_{\parallel} = 2\pi\times66.25$ MHz, so we require $\sqrt{A_{\parallel}^2 + \Omega_e^2} \tau_e = 2m\pi$ for any positive integer $m$ in order for the electron to rotate back to its initial state. This constraint results in discrete choices of the Rabi frequency, namely $\Omega_e = A_{\parallel}/\sqrt{4m^2 -1}$. We choose $m$ to be as small as possible given the constraints on our maximal drive power ($m=2$) to minimize the effect of electron qubit frequency diffusion ($\sim 1$ MHz). For nuclear-flipping pulses, heating limitations force us to work at low Rabi frequencies $\Omega_n \ll \omega_{RF2} - \omega_{RF1}$, so off-resonant transition driving is negligible.

While the above choice of Rabi frequency ensures that the electron is not flipped by off-resonant driving, it still picks up a geometric phase from its rotation on the Bloch sphere, which can be detected when the nucleus is placed in a superposition and C$_n$NOT$_e$ is applied twice:
\[ \ket{\downarrow_e} \left( \ket{\downarrow_n} + \ket{\uparrow_n} \right) \xRightarrow[]{2\times \mathrm{C}_n\mathrm{NOT}_e}  \ket{\downarrow_e} \left( e^{i\gamma_{res}} \ket{\downarrow_n} + e^{4i\gamma_{det}} \ket{\uparrow_n} \right) \] 
where $\gamma_{res}$ and $\gamma_{det}$ refer to the geometric phase picked up by the electron during one closed loop around the resonant and detuned Rabi driving axes respectively, and the factors of 1 and 4 account for the fact that the off-resonant driving causes $4\pi$ rotation for every $\pi$ rotation with resonant driving. The geometric phase can be written as $\gamma = -\frac{S}{2}$ where $S$ is the solid angle on the Bloch sphere subtended by the closed path. In our case, resonant and detuned driving cause the Bloch vector to be driven around the axes $(\Omega_e, 0, 0)$ and  $(\Omega_e, 0, A_{\parallel})$ respectively, which gives the respective geometric phases of $\gamma_{res} = -\pi$ and $\gamma_{det} = -\pi\left( 1-\frac{\sqrt{15}}{4} \right)$. The relative phase $\Gamma$ induced on the nuclear superposition as a result of applying C$_n$NOT$_e$ twice as described above would therefore be $ \Gamma \equiv 4\gamma_{det} - \gamma_{res} \approx 0.87 \pi$.

\begin{figure}[t]
\centering
\includegraphics[width=0.6\textwidth]{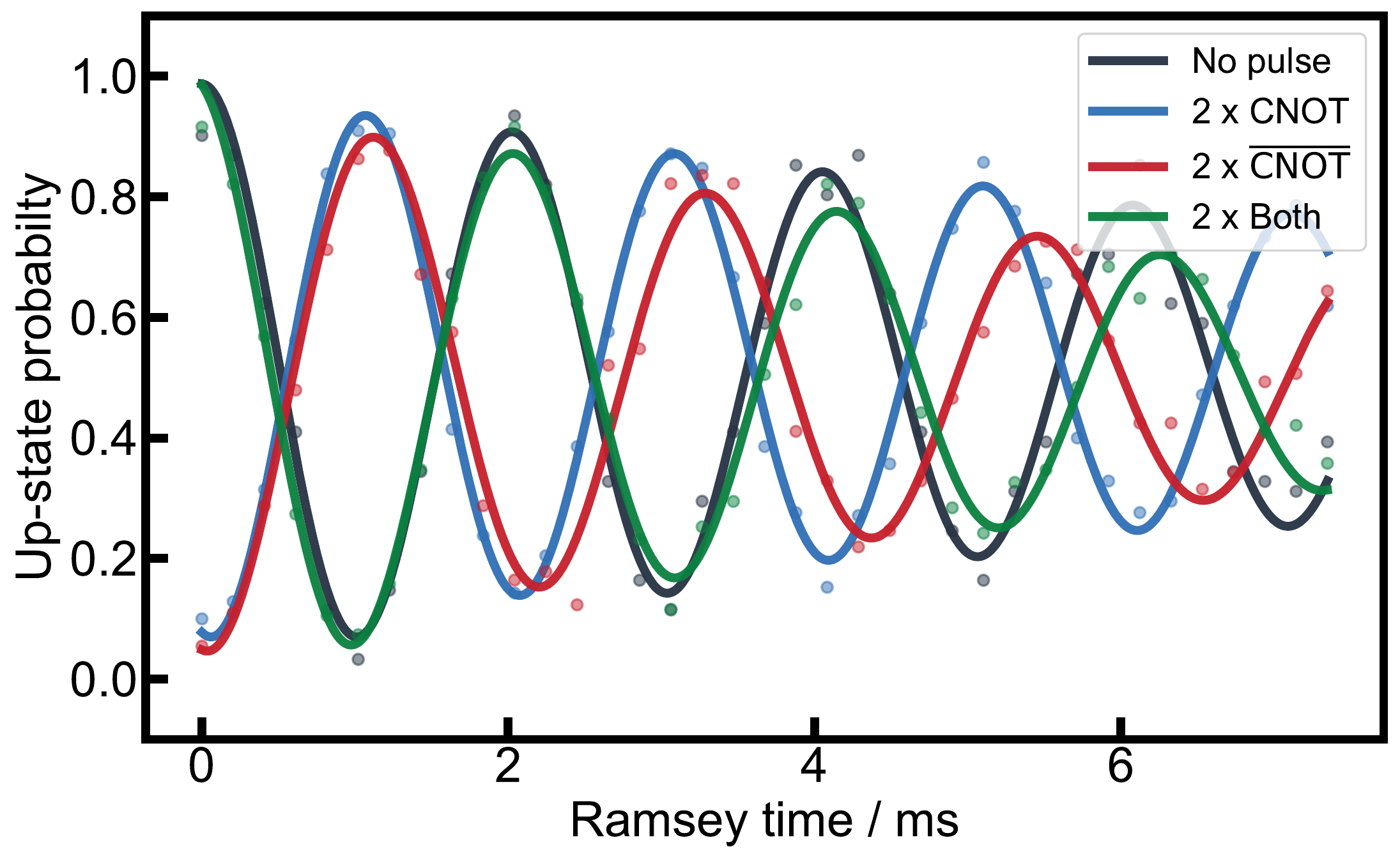}
\caption{Measurement of nuclear Ramsey oscillations with various pulses applied during the free precession time. Nuclear $\pi/2$ pulses were intentionally detuned by 500 Hz so that Ramsey fringes can be observed.}
\label{SIfig:GeomPhase}
\end{figure}

We verify this expression by performing a modified nuclear Ramsey sequence where we drive C$_n$NOT$_e$ pulses on the electron spin during the free evolution time. From Fig. \ref{SIfig:GeomPhase}, we measure that $\Gamma = 0.85 \pi$ and $0.87 \pi$ for C$_n$NOT$_e$ and $\overline{\mathrm{C}_n\mathrm{NOT}_e}$ respectively, which matches our predictions. More importantly, applying both CNOTs twice during the sequence leads to no phase shift, confirming that the geometric phases imparted cancel out. This allows us to neglect this phase as all relevant pulse sequences in our experiments have C$_n$NOT$_e$ and $\overline{\mathrm{C}_n\mathrm{NOT}_e}$ pulses paired appropriately such that this phase is canceled.

For the decoupled C$_e$NOT$_n$, the length of each individual RF pulse was chosen to be precisely an even number of periods. This guarantees that the average magnetic field due to the RF driving in any XY8 waiting window goes to zero, eliminating any extra phase induced on the electron from RF driving. While the pulse sequence for a decoupled $\mathrm{C}_e\mathrm{NOT}_n$ was shown in Fig. \ref{fig:dcpld-CNOT}A of the main text, a decoupled $\overline{\mathrm{C}_e\mathrm{NOT}_n}$ gate can be analogously implemented by switching the positions of the RF1 and RF2 pulses.

\section{\texorpdfstring{\textsuperscript{13}C}{13C} Characterization}

\begin{figure}[ht]
\centering
\includegraphics[trim=0cm 23cm 7cm 0cm]{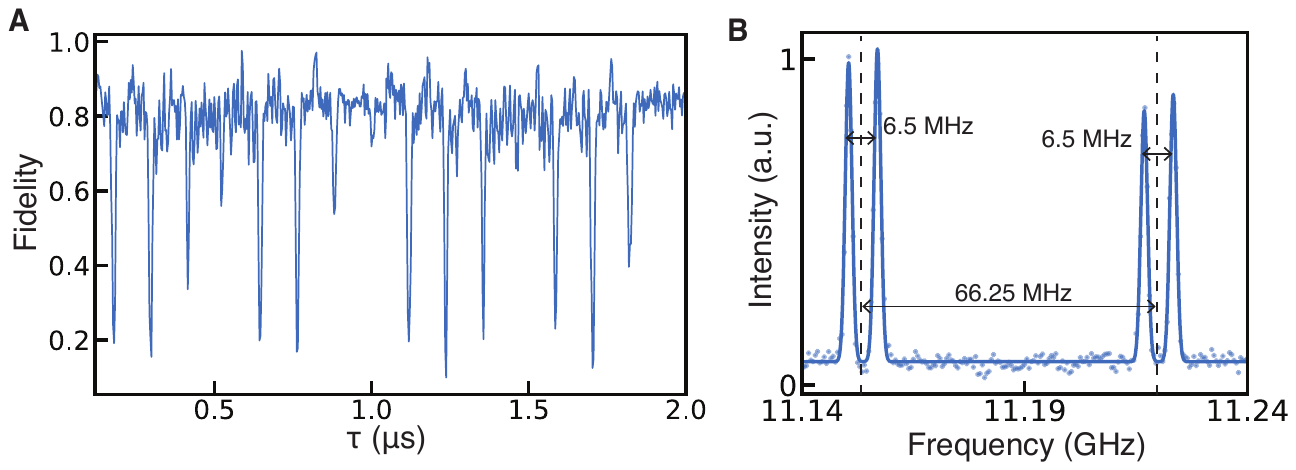}
\caption{(\textbf{A}) XY8 decoupling sequence on the electron, sweeping the inter-pulse time $2\tau$. (\textbf{B}) Optically detected magnetic resonance (ODMR) spectrum of the electron with all nuclear spins uninitialized. The larger splitting (66.25 MHz) is due to the $^{29}$Si spin while the smaller splitting (6.5 MHz) is due to the  $^{13}$C spin.}
\label{SIfig:electron-xy8}
\end{figure}

In addition to the strongly-coupled $^{29}$Si nucleus explored in this paper, our device also contains a comparatively more weakly coupled $^{13}$C nucleus located nearby in the lattice. The coupling is however large enough to induce a splitting of measured electron transitions (Fig. \ref{SIfig:electron-xy8}B). The $^{13}$C nuclear spin was not used in this paper; it is characterized here and was initialized in all of our experiments so that its effect can be neglected. The $^{13}$C-SiV Hamiltonian is similar to the $^{29}$SiV Hamiltonian, though the magnetic field is not necessarily aligned with the hyperfine interaction, resulting in a parallel and perpendicular component of the coupling matrix $\mathbf{A}$~\cite{Smeltzer2011}: 

$$\mathcal{H} = \gamma_e S_z + \gamma_n I_z  + A_{\parallel} S_z I_z + A_{\perp} S_z I_x $$

To evaluate the electron coupling to the nearby $^{13}$C, we performed a sweep of the inter-pulse time of an electron XY8 sequence (Fig. \ref{SIfig:electron-xy8}A). Fitting of this signal indicates a parallel coupling component of $A_{\parallel} = 2\pi\times 3.2$ MHz and a perpendicular coupling of $A_{\perp} = 2\pi\times 0.36$ MHz. The $^{13}$C-flipping RF transitions, which we can drive directly, are 1.041 MHz and 7.475 MHz when the electron spin is in the down and up state respectively. We achieve a Rabi frequency of the transition of 5.23 kHz, corresponding to a $\pi$ time of 95.6 \si{\micro\second}. We use this to initialize the $^{13}$C at the beginning of every experimental sequence by swapping the electron state onto the $^{13}$C (with C$_{{}^{13}\mathrm{C}}$NOT$_e$-C$_e$NOT$_{^{13}\mathrm{C}}$-C$_{^{13}\mathrm{C}}$NOT$_e$), and then confirm whether the initialization was successful by performing a Ramsey sequence on the electron to precisely determine its transition frequency. This allows for a readout of the $^{13}$C state since the $^{13}$C nucleus induces a spin-dependent shift on the electron transition frequency.
\section{Phase Readout of Electron State}

We implement the phase-based readout of the electron state by using an EOM to frequency-shift a carrier laser into two sidebands located 300 MHz above and below the carrier ($\omega_r$ and $\omega_b$, magenta and cyan lines in Fig. \ref{fig:structure}D). Both sideband components travel along the same path, negating the need for any phase stabilization, and are reflected off the device before being detected on a SNSPD with a time resolution of 70 ps (Fig. \ref{fig:structure}E, F). We make use of the fact that there is a sizeable electron state dependence of the reflected phase difference between the two sideband laser frequencies. When these two sidebands are reflected off the cavity, they experience a phase of $\phi^{b}_{\downarrow/\uparrow}$ and $\phi^{r}_{\downarrow/\uparrow}$ respectively, with the phase acquired dependent on the SiV electron state. The value of these phases are given by $\arg(r(\omega_{b}))$ and $\arg(r(\omega_{r}))$, where $r(\omega)$ is the complex reflection amplitude at a laser frequency $\omega$.

\begin{figure}[t]
\centering
\includegraphics[trim=0cm 23cm 6cm 0cm]{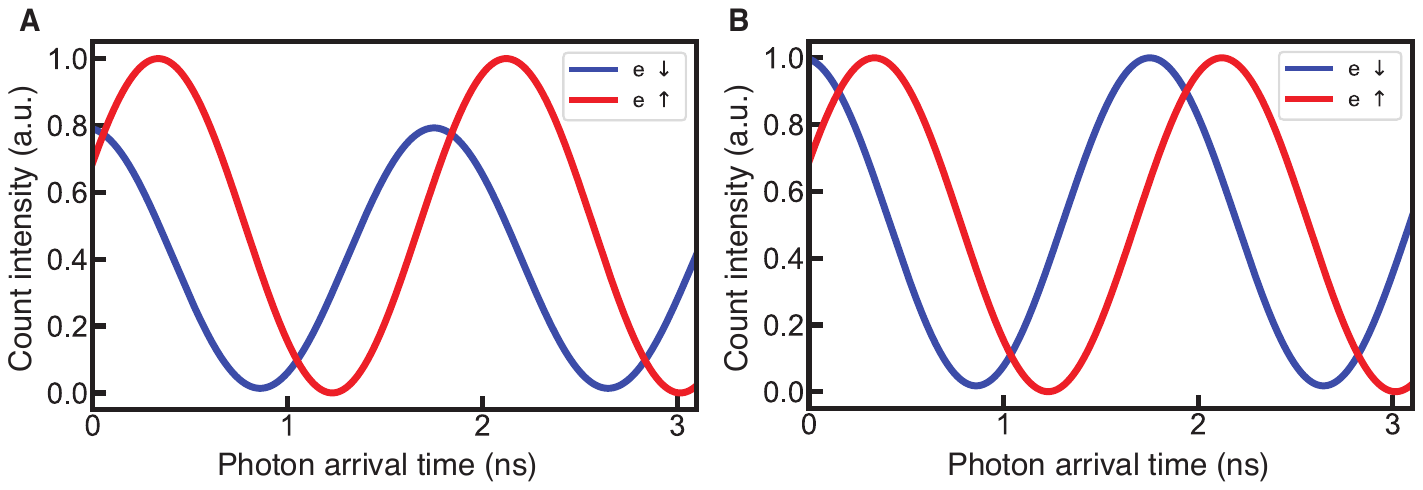}
 \caption{\textbf{(A)} Simulated arrival time probability distributions for both electron states before and \textbf{(B)} after normalization. The normalized distribution as a function of time is called $P_{\downarrow/\uparrow}(t)$. The distributions are shifted in time as the reflected phase difference $\Delta\phi$ between the two measurement sideband frequencies depend on the electron state.}
\label{SIfig:phase-read-hists}
\end{figure}

When these sidebands are sent onto a single photon detector, the photon arrival time probability distribution function is a beat signal at their frequency difference of 600 MHz with a phase of $\Delta\phi_{\downarrow/\uparrow} \equiv (\phi^b_{\downarrow/\uparrow}-\phi^r_{\downarrow/\uparrow})$ depending on the electron spin state. The phase difference for the two electron states $(\Delta\phi_\downarrow - \Delta\phi_\uparrow)$ is observed as a phase shift between the arrival time distributions as seen in Fig. \ref{SIfig:phase-read-hists}A. The experimentally measured beats in Fig. \ref{fig:structure}F are fit using sinusoidal oscillations at 600 MHz with a phase shift computed from the cavity-QED model and the relative amplitudes calculated from the reflectances $|r(\omega_b)|^2$ and $|r(\omega_r)|^2$. A small background signal in the oscillations is modeled as leakage of the carrier frequency through the EOM. 

The fact that the photon arrival time depends on the electron state means that detection of a photon at a given time provides us with information on the electron state, with the information content proportional to the likelihood ratio of the probability distribution functions. We will quantify this information gain using Bayesian analysis and explain how we use this as a readout protocol. We first note that there is an overall scaling difference between the two probability distributions. This is because the spin-up state has a higher reflectance at the phase readout frequencies (Fig. \ref{fig:structure}D), leading to a larger number of counts by 14\%. The curves are separately normalized (Fig. \ref{SIfig:phase-read-hists}B) and we call these the arrival time probability functions $P_{\uparrow/\downarrow}(t)$. 

During the phase readout procedure, the readout lasers are turned on for a fixed period of time and a variable number $N$ of photons are collected that arrive at times $\lbrace t_1, t_2, \dots, t_N\rbrace$. Our initial estimate for the probability that the electron is in spin state $\downarrow$ is chosen to be the uniform prior $p_\downarrow^{(0)}=0.5$. We then repeatedly apply Bayes' rule to update our probability estimate $p_\downarrow^{(i)}$ after the arrival of photon $i$ at time $t_i$. We note that in reality the analysis is done in postprocessing after all $N$ photons have been collected, but it is equivalent to consider performing the analysis in an iterative manner for each individual arriving photon. The posterior probability estimate is:
\begin{equation}
    p_\downarrow^{(i)} = \frac{p_\downarrow^{(i-1)} P_\downarrow(t_i)}{p_\downarrow^{(i-1)} P_\downarrow(t_i) + p_\uparrow^{(i-1)} P_\uparrow(t_i)}
    \label{eqn:bayesArrivalT}
\end{equation}
where $p_\uparrow^{(i)} \equiv 1 - p_\downarrow^{(i)}$. After incorporating the information from the arrival times of all $N$ photons, we apply Bayes' rule one more time, this time using the fact that the two electron states have different overall reflectivities and thus they have a different mean number of reflected photons $\overline{n}_{\uparrow/\downarrow}$ that we expect to detect during the readout window. Our final estimate for the probability that the electron is in the spin-down state is therefore:

\begin{equation}
    p_\downarrow = \frac{p_\downarrow^{(N)} \mathrm{Po}(\overline{n}_{\downarrow}, N)}{p_\downarrow^{(N)} \mathrm{Po}(\overline{n}_{\downarrow}, N) + p_\uparrow^{(N)} \mathrm{Po}(\overline{n}_{\uparrow}, N)}
    \label{eqn:bayesPhotonN}
\end{equation}

Here, $\mathrm{Po}(\overline{n}, N)$ is the probability mass function of a Poisson distribution with mean $\overline{n}$ to have $N$ photons detected. This last factor justifies our decision to separately normalize each electron state's probability distribution previously when defining $P_{\uparrow/\downarrow}(t)$, since the probability of detecting $N$ photons at times  $\lbrace t_1, t_2, \dots, t_N\rbrace$ can be decomposed into (I) the probability of detecting $N$ photons in total (accounted for in Eqn. \ref{eqn:bayesPhotonN}) multiplied by (II) the probability of detecting photons at times  $\lbrace t_1, t_2, \dots, t_N\rbrace$ conditioned on the fact that we have detected $N$ photons in total (accounted for in Eqn. \ref{eqn:bayesArrivalT} by using the normalized distributions). 

At the end of this process, the electron is determined to be in the down state if $p_\downarrow > 0.5$ and in the up state otherwise. In order to reduce the readout error rate, we can perform postselection on the posterior probabilities by determining that the electron is in the down (up) state only if $p_\downarrow > 1 - \varepsilon$ ($p_\downarrow < \varepsilon$) for some stricter choice of threshold $\varepsilon < 0.5$. All readout attempts with $\varepsilon \leq p_\downarrow \leq 1 - \varepsilon$ would be discarded as failed measurements. Such a postselection protocol could be beneficial in scenarios where the accuracy of readout is of paramount importance and where the state preparation can be repeated if necessary (e.g. in a quantum repeater). In the following discussions and in the main text, we fix $\varepsilon=0.2$ for simplicity. 

\begin{figure}[t]
\centering
\includegraphics[width=0.65\textwidth]{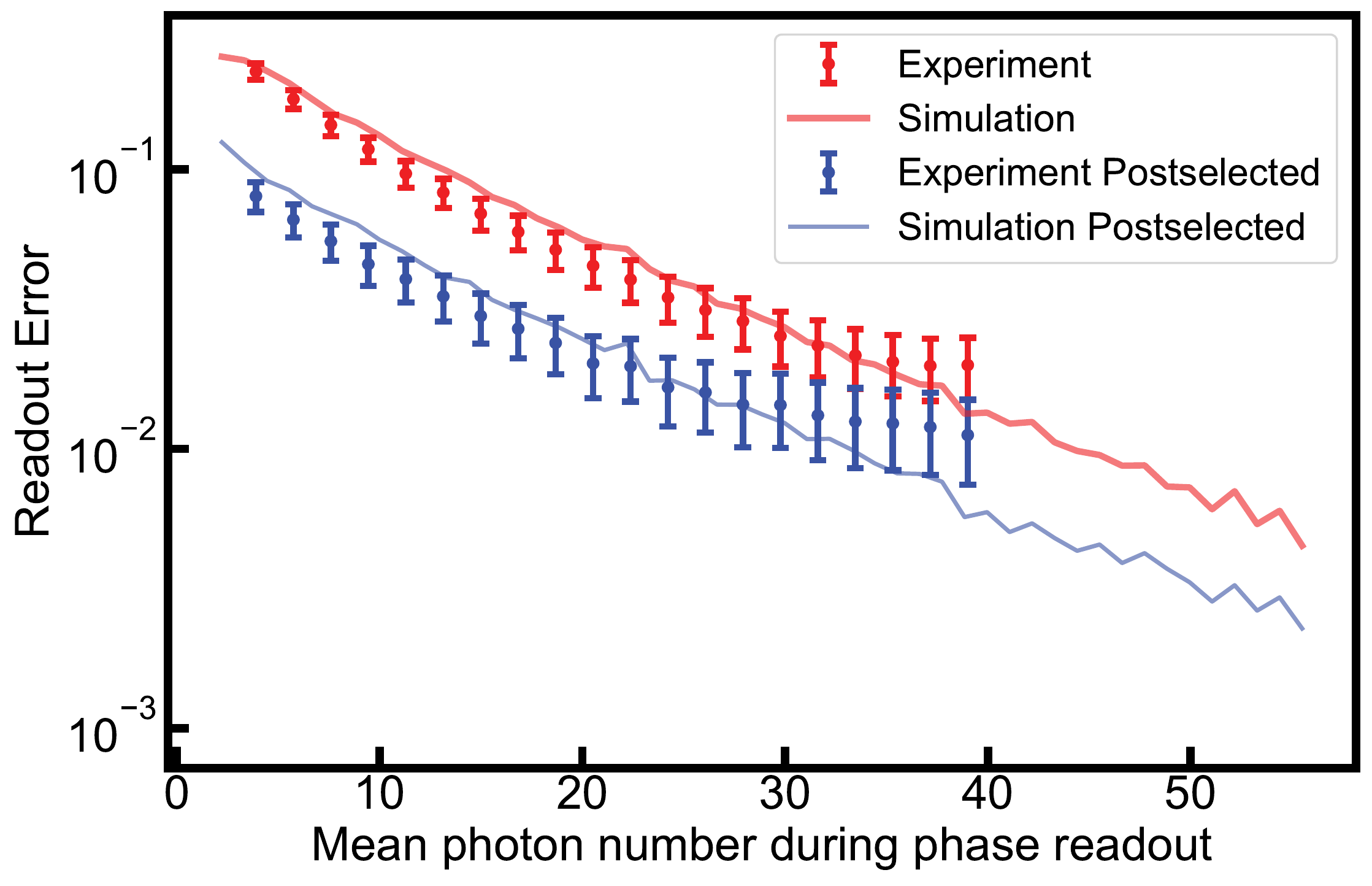}
 \caption{Comparison of phase readout error rates against the mean number of photons collected during each readout window without (red) and with (blue) postselection at 80\% confidence.}
\label{SIfig:phase-read-error-scaling}
\end{figure}

Fig. \ref{SIfig:phase-read-error-scaling} shows how the readout error decreases as we increase the mean number of photons $\overline{n}$ collected during phase readout window by increasing the readout time. The errors are averaged over both spin states. The readout error rate is estimated experimentally by initializing in an $X$-basis state and then read out with the phase readout followed by the resonant readout, with the latter assumed to be the true electron state. 
A simulation of the expected error scaling was performed by generating random photon arrival time samples from the experimentally-measured arrival time distribution.

\begin{figure}[t]
\centering
\includegraphics[width=0.65\textwidth]{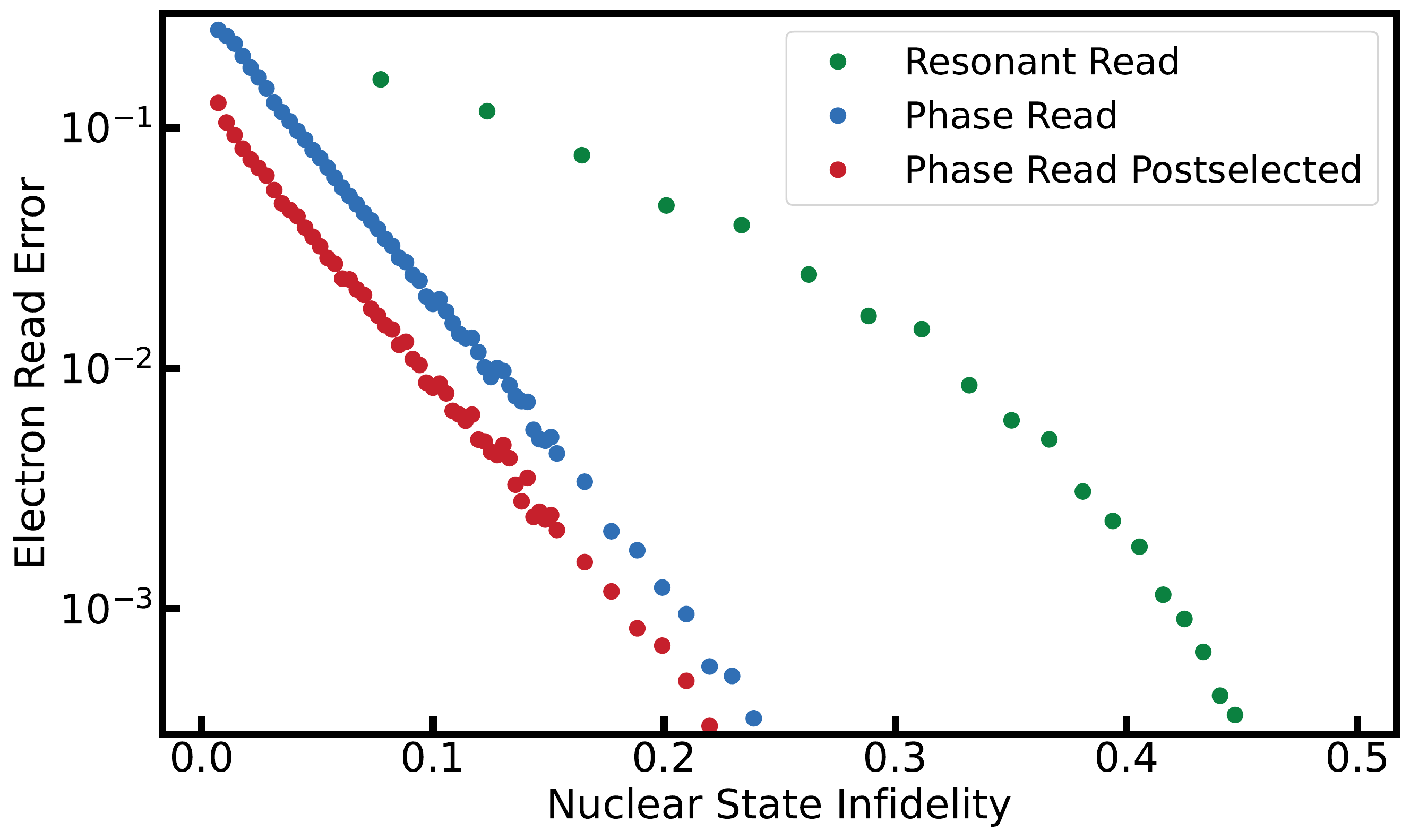}
 \caption{Scaling of simulated electron readout error against nuclear state infidelity caused by the readout. Nuclear state infidelity is defined in a similar manner to Fig. \ref{SIfig:laser-decoherence}, i.e. it is $0$ when we have perfect coherence of a $X$ nuclear superposition state and $0.5$ when the nuclear phase is completely decohered by the electron readout.}
\label{SIfig:phase-read-nuc-scaling}
\end{figure}

This error rate scaling can be combined with our measurement of nuclear decoherence from electron readout (Fig. \ref{SIfig:laser-decoherence}) to characterize the nuclear dephasing rate of the phase readout protocol compared to resonant readout. Fig. \ref{SIfig:phase-read-nuc-scaling} shows that for every desired electron readout fidelity of our choice, the phase read procedure (both with and without postselection) leads to less nuclear state decoherence. For instance, if we desire an electron readout with 5\% error, the electron can be read 14 times using the postselected phase read before we suffer a $1/e$ decoherence of the nuclear state. In contrast, the resonant readout can only be performed 1.8 times before the same amount of nuclear decoherence sets in, which is a factor of 8 lower.

\section{Laser-Induced Nuclear Decoherence}

Due to the difference in the hyperfine coupling between the ground and excited states, the optical transition frequency is shifted depending on the state of the $^{29}$Si nuclear spin, causing the reflection amplitude and phase spectra to be different for different nuclear states (Fig. \ref{fig:structure}D in main text). As a result, the scattered readout photons acquire information about the nuclear state, leading to decoherence of the nuclear superposition once the photon states are traced over. To model the decoherence process, consider a coherent state $\ket{\alpha}$ incident on a uniform superposition of the nuclear state $\frac{1}{\sqrt{2}} \left( \ket{\uparrow} + \ket{\downarrow} \right)$ during a readout step. Here, we assume the electron is in an initialized state and we will omit its state in the following expressions. After scattering the coherent state off the SiV, the light that is reflected, transmitted, and scattered off the cavity can be expressed as:
\[ \ket{\psi_1} = \frac{1}{\sqrt{2}} \left( \ket{\downarrow} \otimes
\ket{R_\downarrow \alpha}\otimes\ket{T_\downarrow \alpha}\otimes\ket{S_\downarrow \alpha} + 
\ket{\uparrow}\otimes\ket{R_\uparrow \alpha}\otimes\ket{T_\uparrow \alpha}\otimes\ket{S_\uparrow \alpha} \right)
 \]
 where $R_{\uparrow/\downarrow}, T_{\uparrow/\downarrow}, S_{\uparrow/\downarrow}$ refer to the reflection, transmission, and scattering coefficients for the cavity with the $^{29}$Si nuclear spin in the state indicated in the subscript. We can then describe the system by the density matrix:
\begin{align*} 
\rho_1 = \ketbra{\psi_1} = \frac{1}{2} (
&\ketbra{\downarrow} \otimes \ketbra{R_\downarrow \alpha}\otimes \ketbra{T_\downarrow \alpha}\otimes \ketbra{S_\downarrow \alpha} +
\\
&\ketbra{\uparrow}\otimes \ketbra{R_\uparrow \alpha}\otimes \ketbra{T_\uparrow \alpha}\otimes \ketbra{S_\uparrow \alpha} +
\\
&\ketbra{\downarrow}{\uparrow}\otimes \ketbra{R_\downarrow \alpha}{R_\uparrow \alpha}\otimes \ketbra{T_\downarrow \alpha}{T_\uparrow \alpha}\otimes \ketbra{S_\downarrow \alpha}{S_\uparrow \alpha} +
\\
&\ketbra{\uparrow}{\downarrow}\otimes \ketbra{R_\uparrow \alpha}{R_\downarrow \alpha}\otimes \ketbra{T_\uparrow \alpha}{T_\downarrow \alpha}\otimes \ketbra{S_\uparrow \alpha}{S_\downarrow \alpha}
)
\end{align*}

The transmitted and scattered light is lost to our experimental setup, while the reflected light is collected and measured by our photon counters (up to collection loss). However, as we do not condition any of the subsequent experimental procedures on the measurement result, we can trace over the reflected photon states as well. Tracing over all the photon states gives the resulting nuclear state:
\begin{align*}
    \rho_1 &= \frac{1}{2} \begin{bmatrix} 
    \braket{R_\downarrow \alpha}{R_\downarrow \alpha}\braket{T_\downarrow \alpha}{T_\downarrow \alpha}\braket{S_\downarrow \alpha}{S_\downarrow \alpha}& 
    \braket{R_\downarrow \alpha}{R_\uparrow \alpha}\braket{T_\downarrow \alpha}{T_\uparrow \alpha}\braket{S_\downarrow \alpha}{S_\uparrow \alpha} \\
    \braket{R_\uparrow \alpha}{R_\downarrow \alpha}\braket{T_\uparrow \alpha}{T_\downarrow \alpha}\braket{S_\uparrow \alpha}{S_\downarrow \alpha} &
    \braket{R_\uparrow \alpha}{R_\uparrow \alpha}\braket{T_\uparrow \alpha}{T_\uparrow \alpha}\braket{S_\uparrow \alpha}{S_\uparrow \alpha}
    \end{bmatrix} \\ &\equiv \frac{1}{2}
    \begin{bmatrix} 1 &  \rho_{1,\downarrow\uparrow} \\
    \rho_{1,\downarrow\uparrow}^* & 1
    \end{bmatrix}  
\end{align*}
where the matrix is written in the nuclear state basis $\lbrace \ket{\downarrow}, \ket{\uparrow} \rbrace$. The off-diagonal term $\rho_{1,\downarrow\uparrow}$ measures the coherence of the state and is given by:
\[ \rho_{1,\downarrow\uparrow} = \prod_{C\in\lbrace R,T,S \rbrace} \braket{C_\downarrow\alpha}{C_\uparrow\alpha} = \prod_{C\in\lbrace R,T,S \rbrace} \exp\left[ -\frac{1}{2}\left( |C_\downarrow \alpha|^2 + |C_\uparrow \alpha|^2\right) + (C_\downarrow \alpha)^*(C_\uparrow \alpha) \right]
\] 
where $C$ runs over all three paths of the light and the final term arises from the inner product properties of coherent states. The term in the product can be reformulated as:
\[
\exp\left[ -\frac{1}{2}\left( |C_\downarrow \alpha - C_\uparrow \alpha|^2\right) + \frac{1}{2}\left( (C_\downarrow \alpha)^*(C_\uparrow \alpha) - (C_\downarrow \alpha)(C_\uparrow \alpha)^* \right) \right]
\]
We can add further clarity by noting that $a^*b-ab^*$ is purely imaginary for all complex numbers $a$ and $b$ since it can be rewritten as $2i \left( \Re(a)\Im(b) - \Im(a)\Re(b) \right)$, and thus we can rewrite $(C_\downarrow \alpha)^*(C_\uparrow \alpha) - (C_\downarrow \alpha)(C_\uparrow \alpha)^*$ as $2i\phi$ for some real $\phi$ that depends on $\alpha$. This allows us to simplify:
\[ \rho_{1,\downarrow\uparrow} = \prod_{C\in\lbrace R,T,S \rbrace} \exp\left[ -\frac{1}{2}\left( |C_\downarrow \alpha - C_\uparrow \alpha|^2\right) \right] e^{i\phi} \]
separating the effect of the light into a pure loss of coherence and a pure rotation similar to a Stark shift.

The coherence can then be measured by measuring the expectation value of the nuclear state in the $X$-basis, which gives:
\[ \langle{X}_1\rangle = \Tr X\rho_1 =  \frac{\rho_{1,\downarrow\uparrow}+\rho_{1,\downarrow\uparrow}^*}{2} = \Re \rho_{1,\downarrow\uparrow} \]
This is the experimental quantity that we measure to determine the decoherence induced on the nucleus by scattering photons.

We can cancel out the coherent phase rotation on the nucleus by performing a $\pi$ phase flip on the nucleus between two separate coherent state packets $\ket{\alpha}$, in a manner similar to that in a spin echo experiment. This effectively performs the state readout in two separate windows. The resulting state can be derived using an almost identical process, giving: 
\begin{align*} |\psi_2\rangle = \frac{1}{\sqrt{2}} ( & \ket{\uparrow} \otimes
\ket{R_\downarrow \alpha}\otimes\ket{T_\downarrow \alpha}\otimes\ket{S_\downarrow \alpha} \otimes\ket{R_\uparrow \alpha}\otimes\ket{T_\uparrow \alpha}\otimes\ket{S_\uparrow \alpha} + \\
&\ket{\downarrow}\otimes\ket{R_\uparrow \alpha}\otimes\ket{T_\uparrow \alpha}\otimes\ket{S_\uparrow \alpha} \otimes
\ket{R_\downarrow \alpha}\otimes\ket{T_\downarrow \alpha}\otimes\ket{S_\downarrow \alpha}  )
 \end{align*}
where the first (second) three photonic kets refer to photons scattered in the first (second) time window. Taking the density matrix as before:
\[ \rho_2 = \frac{1}{2}
\begin{bmatrix} 1 &  \rho_{2,\downarrow\uparrow} \\
\rho_{2,\downarrow\uparrow}^* & 1
\end{bmatrix}  \]
with
\[ \rho_{2,\downarrow\uparrow} = \prod_{C\in\lbrace R,T,S \rbrace} \left\vert \braket{C_\downarrow\alpha}{C_\uparrow\alpha} \right\vert ^2 = \prod_{C\in\lbrace R,T,S \rbrace} \exp\left[ -\left\vert C_\downarrow\alpha - C_\uparrow\alpha \right\vert ^2 \right]
\]
which gives rise to a pure exponential decay with the oscillations canceled out. As before, the experimental observable is the $X$ expectation value which is given by:
\[ \langle{X}_2\rangle = \Tr X\rho_2 =  \Re \rho_{2,\downarrow\uparrow} = \rho_{2,\downarrow\uparrow} \]

All desired dependencies of the nuclear decoherence rate on the readout parameters can be derived from these expressions for the $X$ expectation value. In particular, the laser readout frequency affects the reflection, transmission and scattering coefficients $R, T, S$ according to the Jaynes-Cummings model, while the readout length and laser intensity affects the coherence state $\alpha$ via the mean photon number $N$ in each readout packet with $\langle N\rangle = |\alpha|^2$.

\begin{figure}[t]
\centering
\includegraphics[width=13cm, trim=0cm 24cm 7cm 0cm]{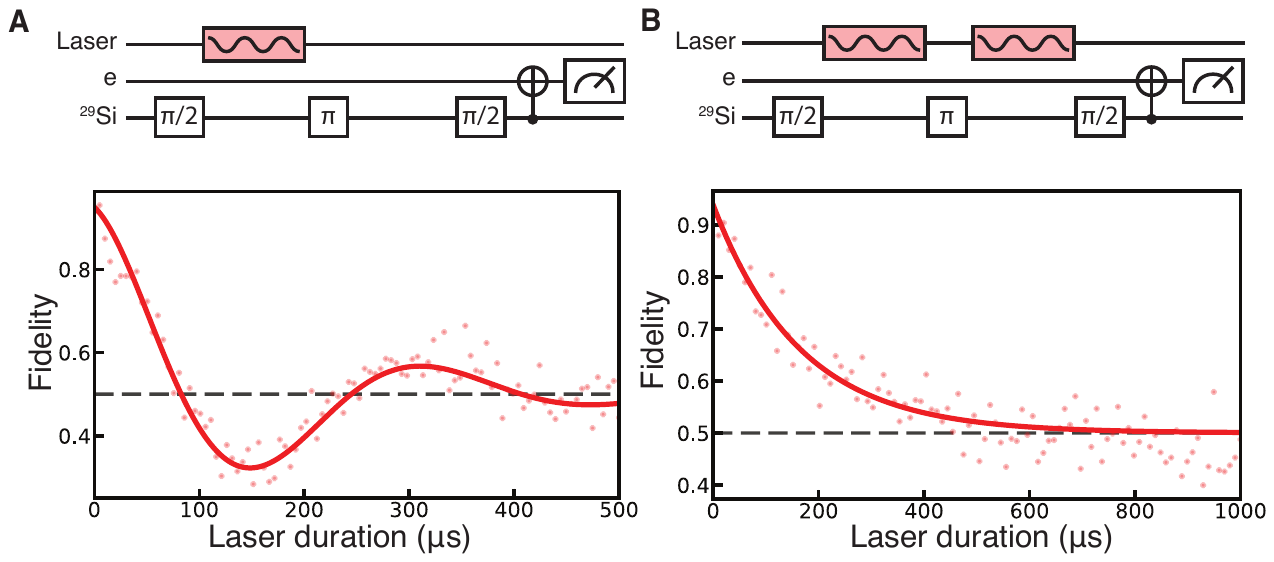}
 \caption{\textbf{(A)} Comparison of nuclear decoherence rate against laser duration when a laser at the resonant readout frequency is applied in an asymmetric and \textbf{(B)} symmetric manner during a spin echo measurement. The resulting coherence of the nucleus is read out in the $X$ basis by applying a $\pi/2$ pulse.}
\label{SIfig:laser-decoherence}
\end{figure}

\begin{figure}[t]
\centering
\includegraphics[width=12cm, trim=0cm 22.25cm 10cm 0cm]{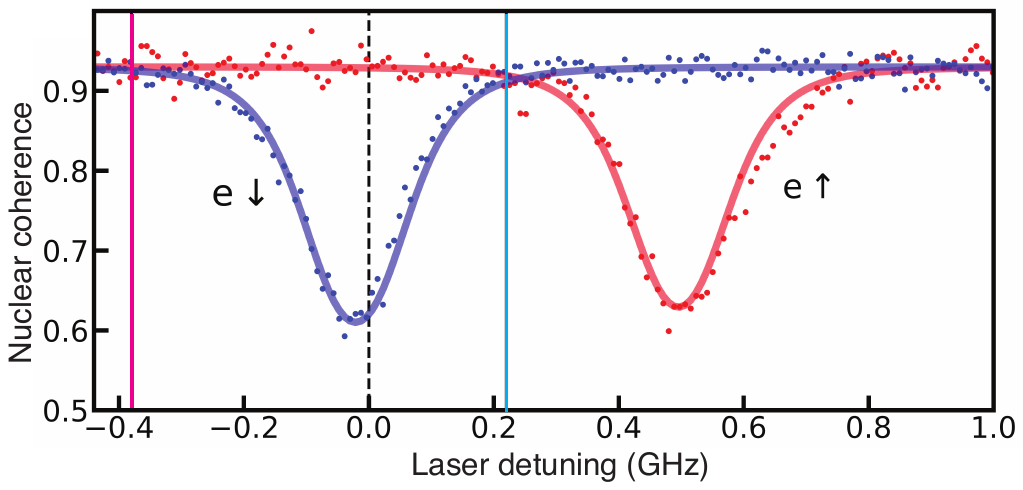}
\caption{Nuclear decoherence from symmetric application of a laser during spin echo process as  laser frequency is varied. The amount of nuclear decoherence experienced depends on the electron state. The dashed vertical line is the laser position for resonant readout; the cyan and magenta vertical lines are the frequencies for phase readout.}
\label{SIfig:laser-decoherence-vary-freq}
\end{figure}

We verify this model of decoherence with the experimental sequence shown in Fig. \ref{SIfig:laser-decoherence}. The $^{29}$Si nuclear spin is first prepared in a superposition with a $\frac{\pi}{2}$ pulse, and the readout laser is applied in either one or two readout windows separated by a $\pi$ pulse on the nucleus. The nuclear spin is then read out in the $X$ basis by applying a $\frac{\pi}{2}$ pulse followed by reading out via the electron. Fig. \ref{SIfig:laser-decoherence}A shows that the nuclear coherence exhibits oscillations from the coherent accumulation of phase and an exponential decay from the decoherence process predicted above, while applying the laser pulse in a symmetrical manner around the $\pi$ pulse cancels out the oscillation and highlights only the laser-induced decoherence (Fig. \ref{SIfig:laser-decoherence}B). The plots are fit with the expressions derived for $\langle X_1\rangle$ and $\langle X_2 \rangle$ respectively using the reflection, transmission, and scattering coefficients computed from the cavity-QED model.

This model can be further explored by sweeping the frequency of the laser being applied while keeping the laser length constant (Fig. \ref{SIfig:laser-decoherence-vary-freq}). The amount of decoherence experienced by the nucleus at a given laser frequency depends on the electron state. The black dashed line indicates the resonant readout frequency, i.e. where the SiV contrast is largest, and is also the frequency used above in Fig. \ref{SIfig:laser-decoherence}. The fact that it lies so close to a minimum of the nuclear coherence confirms the argument in the main text that resonant readout, while being a high-fidelity electron state readout method, also reads out the nuclear state and leads to rapid decoherence of any nuclear superposition. In contrast, the frequencies used for the phase readout (cyan and magenta lines) lead to much weaker nuclear decoherence as they do not reveal much information about the nuclear state in both amplitude and phase (Fig. \ref{fig:structure}D), making them suitable for repeatedly reading and resetting the electron while using the nucleus as a long-lived memory.

\section{Gate fidelity decomposition}

\begin{figure}[t]
\centering
\includegraphics[trim=0cm 21.5cm 11cm  0cm]{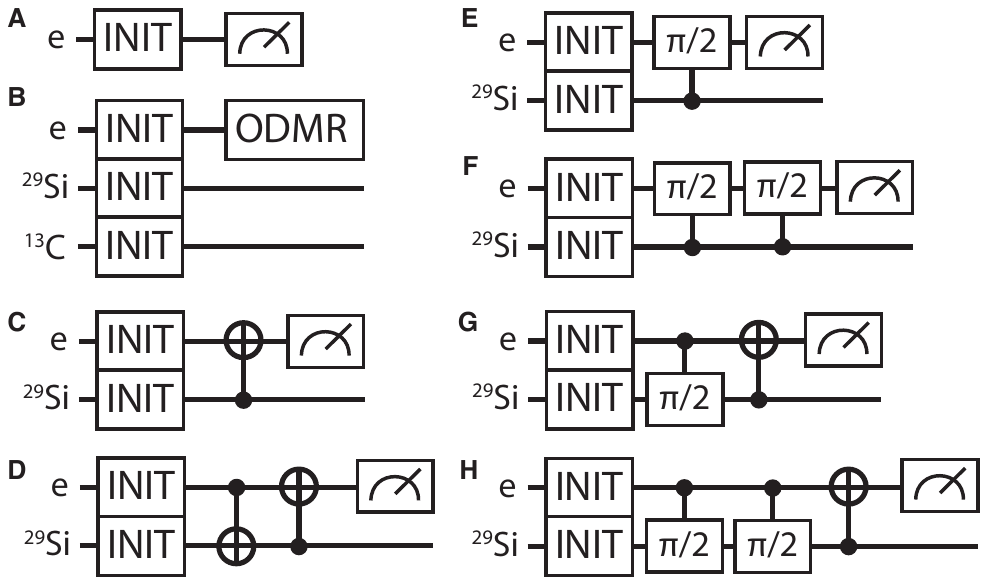}
\caption{(\textbf{A}) through (\textbf{H}): sequences to evaluate the fidelities of individual sequence components as described in the text. All initialization steps set the spins into the spin-down states, and all CNOT gates are conditioned on the control spin being in the spin-down state.}
\label{SIfig:fid_seq}
\end{figure}

Here, we characterize the fidelities of the various gates and constituent operations in our system. Initialization of the electron spin in the spin-down (spin-up) state is performed by applying a resonant readout laser pulse and checking if the reflected photon counts are below (above) a set initialization threshold. If not, a $\pi$ pulse is applied to the electron and the readout is repeated until the desired state is measured. The initialization threshold for initializing in the spin-down (spin-up) state is lower (higher) than the readout threshold used for state determination, which allows for higher initialization fidelity and mitigates spectral diffusion by only allowing the initialization to succeed when there is sufficient reflection contrast between the spin states. By initializing the electron the down state and subsequently reading it out, we determine that our combined initialization and readout fidelity in the down state is $99.5\pm0.1\%$ (Fig. \ref{SIfig:fid_seq}A). Readout of the nuclear state is performed with a CNOT followed by resonant electron readout (Fig. \ref{SIfig:fid_seq}C).

To initialize both the $^{29}$Si and $^{13}$C nuclear spins, we initialize the electron in the spin-down state and swap the state onto the $^{29}$Si before repeating the same steps for the $^{13}$C. To confirm the success of the swap gates, we perform a Ramsey sequence on the lowest-frequency electron transition (corresponding to the $^{29}$Si$\downarrow$ $^{13}$C$\downarrow$ state) to determine if the entire system is in the correct state (e$\downarrow$ $^{29}$Si$\downarrow$ $^{13}$C$\downarrow$), where successful initialization is indicated by the lack of rapid Ramsey fringes that would be observed if the system were in a different state (which would have a detuned transition frequency). We then do an electron ODMR sequence to determine the nuclear populations and find that the $^{29}$Si is correctly initialized $99.5\pm0.1\%$ of the time and the $^{13}$C is correctly initialized $93.8\pm1.5\%$ of the time (Fig. \ref{SIfig:fid_seq}B).

We can now characterize the fidelities of the various qubit operations using the circuits as shown in Figs. \ref{SIfig:fid_seq}C-H and we summarize the results in Table \ref{tab:fid-seqc-results}. These gate fidelities can be combined with the overall fidelity of the swap-hold-read sequence as performed in Fig. \ref{fig:dcpld-CNOT}D to find that the decoupled $\mathrm{C}_e\mathrm{NOT}_n$ has a fidelity of $93.7\pm0.7\%$ (Table \ref{tab:fidelities}). The fidelity loss is mainly due to electron decoherence during the decoupling, as well as cumulated $\mathrm{C}_n\mathrm{NOT}_e$ and $\overline{\mathrm{C}_n\mathrm{NOT}_e}$ gate infidelities.

\begin{table}
    \centering
    \begin{tabular}{ccc}
        \hline
        Sequence & P($\uparrow$) & Deduced Gate Fidelity \\ 
        \hline \hline
        C & $99.4\pm0.1\%$ & $\mathrm{C}_n\mathrm{NOT}_e$ fidelity: $0.999\pm0.001$ \\
        \hline
        D & $2.5\pm0.2\%$  & $\mathrm{C}_e\mathrm{NOT}_n$ fidelity: $0.980\pm0.002$ \\
        \hline
        E & $50.6\pm0.6\%$ & \multirow{2}{*}{$\sqrt{\mathrm{C}_n\mathrm{NOT}_e}$ fidelity: $0.988\pm0.002$} \\
        \cline{0-1}
        F & $97.1\pm0.2\%$ & \\
        \hline
        G & $49.7\pm0.6\%$ & \multirow{2}{*}{$\sqrt{\mathrm{C}_e\mathrm{NOT}_n}$ fidelity: $0.990\pm0.002$} \\
        \cline{0-1}
        H & $98.0\pm0.2\%$ \\
        \hline
    \end{tabular}
    \caption{Outcomes of sequences shown in Fig. \ref{SIfig:fid_seq} with corresponding deduced gate fidelities.}
    \label{tab:fid-seqc-results}
\end{table}

\begin{table}
\centering
\begin{tabular}{cc} 
 \hline
 Step & Fidelity \\ 
 \hline\hline
 Initialization & $0.995\pm0.001$ \\
 \hline
 $\sqrt{\mathrm{C}_n\mathrm{NOT}_e}$ & $0.988\pm0.002$ \\ 
 \hline
 \textbf{Decoupled} $\mathbf{C}_{\mathbf{e}}\mathbf{NOT}_{\mathbf{n}}$ & \textbf{$0.937\pm0.007$} \\
 \hline
 $\mathrm{C}_n\mathrm{NOT}_e$ & $0.999\pm0.001$ \\
 \hline
 $\mathrm{C}_e\mathrm{NOT}_n$ & $0.980 \pm0.002$ \\
 \hline
 $\sqrt{\mathrm{C}_e\mathrm{NOT}_n}$ & $0.990\pm0.002$ \\ 
 \hline
 $\mathrm{C}_n\mathrm{NOT}_e$ & $0.999\pm0.001$ \\
 \hline
 Readout & $0.995\pm0.001$ \\
 \hline\hline
 Total measured fidelity & $0.891\pm0.006$ \\
 \hline
 Total estimated fidelity & $0.891\pm0.008$ \\
 \hline
 \end{tabular}
\caption{\label{tab:fidelities} Fidelity decomposition for the Swap-Hold-Read sequence. The predicted fidelity is from the product of the fidelities of all steps.}
\end{table}

\section{Spin-photon gates operation and measurement}
The initial photonic state for all spin-photon entangling gates is prepared as weak coherent pulses using an EOM. To limit the occurrence of two-photon events that would lower the fidelity of the final spin-photon entangled state, we work with pulses with average photon number $5 \times 10^{-3}$. The early and late photon timebins are separated by $\delta t = \SI{143.5}{\nano\second}$ and have a Gaussian pulse shape of width \SI{20.8}{\nano\second}. The pulse width is chosen to be as wide as possible while still fully fitting within the separation time together with the microwave gates. This minimizes the spectral width of the photons, resulting in a better optical contrast. For the 4.3K operation of the PHONE gate, the photon pulses are shortened to square pulses of width \SI{16.7}{\nano\second} so that the C$_n$NOT$_e$ and $\overline{\mathrm{C}_n\mathrm{NOT}_e}$ pulse pairs can be placed closer together. This is done to limit decoherence from the electron by lowering the time it is in a superposition state.

Measurement of the final entangled state is done by individually measuring the spin and the photon in different bases. To measure the photon in the $Z$ basis, the photon arrival time is recorded to determine if it is in the early or late timebin. To measure the photon in the $X$ and $Y$ bases, we use an in-fiber TDI with a delay line equal to the photon spacing 143.5 ns corresponding to a fiber length of \SI{43.05}{\meter}. The exact length of the delay line is adjusted with a piezoelectric hollow cylinder around which the fiber is wrapped. The TDI is locked every $\sim$5 s by sending CW locking light through the locking port (Fig. \ref{SIfig:setup}A) and adjusting the delay line length to equalize counts on both outputs of the TDI. To measure in a specific basis, the frequency difference between the locking light and the photonic qubits is adjusted by changing the MW frequency sent to the locking light AOM (effectively changing the lock point) so that maximum photonic qubit contrast is achieved at the output, i.e. $\ket{+X}$ (or $\ket{+Y}$) results in clicks at APD1 only, and $\ket{-X}$ (or $\ket{-Y}$) results in clicks at APD2 only.

\begin{figure}[t]
\centering
\includegraphics[trim=0cm 24cm 14cm  0cm]{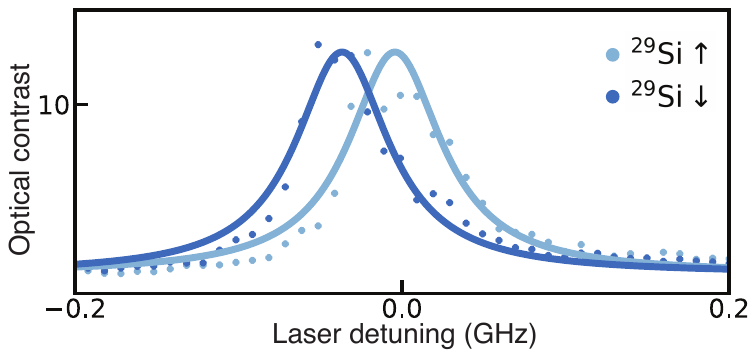}
\caption{Frequency-dependent contrast between the electron spin states for the two nuclear spin states.}
\label{SIfig:contrast_nuc}
\end{figure}

The measurements are combined to calculate the overlap fidelity with the $\ket{\Phi^+} = (\ket{e\downarrow} + \ket{l\uparrow})/\sqrt{2} \equiv (\ket{00} + \ket{11})/\sqrt{2}$ Bell state. For simplicity, here we denote the $\ket{e}$ and $\ket{\downarrow}$ states by $\ket{0}$ and the $\ket{l}$ and $\ket{\uparrow}$ states by $\ket{1}$, where the spin states can correspond to either to the electron or nuclear spin depending on the entangled state that we are preparing. The overlap fidelity with a given two-qubit density matrix $\rho$ is given by: 
$$F = \bra{\Phi^+}\rho\ket{\Phi^+} = \frac{1}{2}(\rho_{11} + \rho_{44} + 2\Re (\rho_{14}))$$
We want to express the above density matrix entries in terms of experimental observables. When the two-qubit system is measured in the $XX$, $YY$, and $ZZ$ bases, the probability of obtaining the various measurement outcomes $(i,j) \in \{ 0,1 \}^2$ are:
$$p_{zz}^{ij} = \bra{ij}\rho\ket{ij}$$
$$p_{xx}^{ij} = \bra{ij}(\sqrt{X}_{\mathrm{spin}}\otimes\sqrt{X}_{\mathrm{photon}})\rho(\sqrt{X}_{\mathrm{spin}}\otimes\sqrt{X}_{\mathrm{photon}})\ket{ij}$$
$$p_{yy}^{ij} = \bra{ij}(\sqrt{Y}_{\mathrm{spin}}\otimes\sqrt{Y}_{\mathrm{photon}})\rho(\sqrt{Y}_{\mathrm{spin}}\otimes\sqrt{Y}_{\mathrm{photon}})\ket{ij}$$
where the operators $X$ and $Y$ describe a $\pi$-rotation around the $x$ and $y$-axes respectively: $\sqrt{X}=\exp{i\pi S_x /2\hbar}$, $\sqrt{Y}=\exp{i\pi S_y /2\hbar}$ with Pauli spin-\nicefrac{1}{2} matrices $S_x$ and $S_y$. The measurement outcome probabilities can then be combined into the following linear combinations that relate to the relevant entries in the density matrix:
\begin{gather*} 
    P_{zz} \equiv p_{zz}^{00} + p_{zz}^{11} = \rho_{11} + \rho_{44} \\
    P_{xx} \equiv p_{xx}^{00} + p_{xx}^{11} - p_{xx}^{01} - p_{xx}^{10} = 4\Re (\rho_{14}) + 4\Re (\rho_{23}) \\
    P_{yy} \equiv p_{yy}^{01} + p_{yy}^{10} - p_{yy}^{00} - p_{yy}^{11} = 4\Re (\rho_{14}) - 4\Re (\rho_{23})
\end{gather*}
so that we are finally left with an explicit definition of the fidelity in terms of our measurement outcomes:
$$F = \frac{1}{2}P_{zz} + \frac{1}{4}P_{xx} + \frac{1}{4}P_{yy}$$
A decomposition of the sources of errors for spin-photon gates is given in Table \ref{tab:spin-photon-fidelities}. Note that the difference in SiV contrast between the electron-photon and PHONE gates is due to the nuclear state-dependence of the optical transition frequency (Fig. \ref{fig:structure}D) and thus contrast (Fig. \ref{SIfig:contrast_nuc}). This means that the maximum achievable optical contrast for the PHONE gate will be lowered when we average over both nuclear spin states.

\begin{table}
\centering
\begin{tabular}{ccccc} 
 \hline
  & \multicolumn{2}{c}{Electron-photon} & \multicolumn{2}{c}{PHONE} \\ 
 \hline\hline
 Measurement Basis & $ZZ$ & $XX, YY$ & $ZZ$ & $XX, YY$ \\ 
 \hline \hline
 Temperature & \multicolumn{2}{c}{0.1 K (1.5 K)} & \multicolumn{2}{c}{0.1 K (4.3 K)} \\
 \hline
 Initialization & \multicolumn{2}{c}{0.995} & \multicolumn{2}{c}{0.995 (0.970)} \\
 \hline
 MW gates & \multicolumn{2}{c}{0.99} & \multicolumn{2}{c}{0.94 (0.89)} \\ 
 \hline
 SiV contrast & \multicolumn{2}{c}{0.96} & \multicolumn{2}{c}{0.94} \\
 \hline
 Dark counts & \multicolumn{2}{c}{0.997} & \multicolumn{2}{c}{0.997} \\
 \hline
 2-photon events & \multicolumn{2}{c}{0.998} & \multicolumn{2}{c}{0.998} \\
 \hline
 Readout & \multicolumn{2}{c}{0.995} & \multicolumn{2}{c}{0.995} \\ 
 \hline
 TDI contrast & N/A & 0.967 & N/A & 0.967 \\
 \hline
 TDI lock and polarization drift & N/A & 0.98 & N/A & 0.98 \\
 \hline
 $T_2$ dephasing & N/A & 0.997 & N/A & 0.997 (0.680) \\
 \hline\hline
 Total estimated fidelity & \multicolumn{2}{c}{0.91 (0.91)} & \multicolumn{2}{c}{0.85 (0.66)} \\ 
 \hline
 Total measured fidelity & \multicolumn{2}{c}{$0.91\pm0.02$ ($0.90\pm0.01$)} & \multicolumn{2}{c}{$0.85\pm0.02$ ($0.66\pm0.02$)} \\
 \hline
 \end{tabular}
\caption{\label{tab:spin-photon-fidelities} Fidelity decomposition for all spin-photon gates. Parameters that take on a different value at temperatures of 1.5 K and 4.3 K for the electron-photon and PHONE gates respectively are shown in parentheses.}
\end{table}

To determine the spin-photon heralding efficiency, we perform a spin echo sequence on the SiV electron. During the echo wait time, we apply a laser pulse of variable duration. We collect the reflected photons from this laser pulse and measure the state of the electron at the end of the echo sequence. Similar to the effect on the nucleus, reading out the electron during a spin echo dephases the superposition and causes the final electron state to exponentially decay towards a completely mixed state as a function of the average photon number arriving at the SiV. This gives us an estimate for $N$, the average number of photons that have interacted with the SiV, as a function of laser pulse time. 

On the other hand, the number of reflected photons collected $N_c$ during the sequence increases linearly with $N$. By plotting $N_c$ against $N$ for various laser pulse times, we obtain an estimate for the slope $\eta$, which represents the efficiency with which photons that have interacted with the SiV are able to be collected in the photodetectors, which is also the spin-photon heralding efficiency. Since this measurement was performed with SNSPDs with quantum efficiency near $\sim$1 \cite{Bhaskar2020}, we neglect this as a source of loss. We find that the efficiency $\eta = 0.143$. A major source of loss is the low reflectivity of the SiV  ($\sim$0.275, averaged over both electron states), while the remainder can be attributed to the coupling efficiency of the tapered fiber from the nanophotonic device ($\sim$0.6 -- 0.7), as well as losses in splicing points and in the 99:1 in-fiber beamsplitter along the path.

\section{Temperature Robustness with SiV Strain}

Within our experimental temperature regime of 100 mK to 4 K, SiVs experience temperature-dependent spin decoherence \cite{Meesala2018,Pingault2017} primarily through a resonant two-phonon process that involves the orbital upper-branch ground states as an intermediate state. Electrons from the qubit manifold are excited by thermal phonons to the upper orbital states for a random amount of time before decaying via another phonon back to the lower orbital states. In this process, the electron spin state is preserved as our magnetic fields are aligned with the SiV symmetry axis, but the different precession rate in the upper branch means that the electron obtains a random phase and becomes dephased after many such stochastic excitations. The rate of this two-phonon process is proportional to the Bose-Einstein distribution of phonon occupancy at the ground state splitting energy $\Delta_{GS}$ scaling as $(e^{h\Delta_{GS}/k_BT}-1)^{-1}$, so we expect that SiVs with a larger $\Delta_{GS}$ will have coherence times that are less sensitive to increasing temperature \cite{Jahnke2015,Meesala2018}. There is additionally a weaker decoherence effect from the direct single-phonon spin-flipping process that limits electron $T_2$ through electron depolarization, but the rate of this process is lower as it scales with the relatively smaller qubit frequency $\omega_{\mathrm{qubit}}\sim\SI{12}{GHz}$ as $\omega_{\mathrm{qubit}}^3$ \cite{Meesala2018}. The dominant limitation on the value of $T_2$ for our SiV electron spins is therefore two-phonon processes to the upper ground state branches. 

To verify that the sustained long coherence times of our device at high temperatures (Fig. \ref{fig:spin-photon-gates}A) is indeed due to its large $\Delta_{GS}$ suppressing the rate of two-phonon events, we can look at a closely related quantity, the electron spin relaxation rate $T_1^{-1}$. The spin lifetime is also limited by the two-phonon process \cite{Jahnke2015,Meesala2018} since the electron has a small chance of undergoing a spin-flipping decay from the upper orbital state when the magnetic field is not perfectly aligned \cite{Sukachev2017}; this is also known as the Orbach process \cite{Orbach1961}. The direct single-phonon spin-flipping process also has a contribution to $T_1^{-1}$ that is relevant at lower temperatures ($k_BT/\hbar\ll\omega_{\mathrm{qubit}}$), while the two-phonon process dominates at higher temperatures comparable to the ground state splitting ($k_BT/\hbar \sim \Delta_{GS}$) \cite{Jahnke2015,Meesala2018}. The crossover point between these two mechanisms with increasing temperature will be observable as an inflection with a sharp decrease in spin lifetime and coherence. 

\begin{figure}[t]
\centering
\includegraphics[trim=0cm 22.25cm 13cm 0cm]{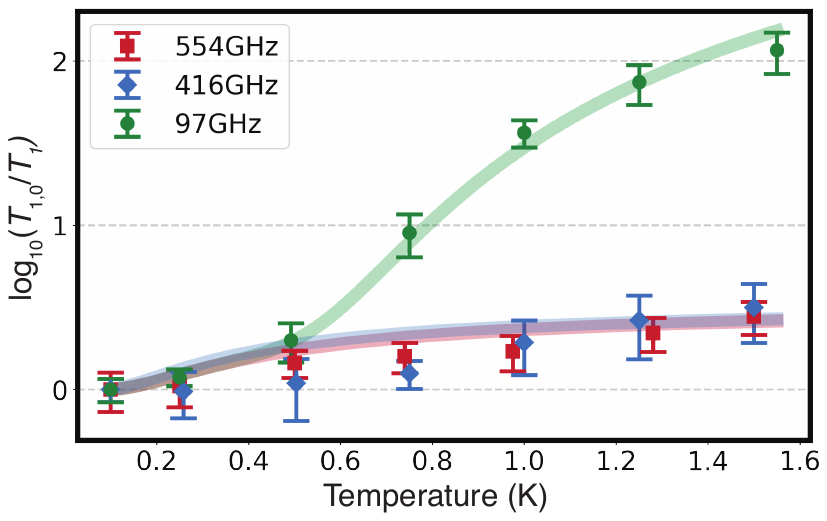}
\caption{Comparison of electron $T_1^{-1}$ (normalized to $T_{1,0}$, their value at 100 mK) as a function of temperature for various SiVs in the same device with different ground state splittings $\Delta_{GS}$. The ground state splittings $\Delta_{GS}$ for all SiVs were measured at zero magnetic field at 100 mK by driving resonantly on the spin-conserving $C$ transition and collecting fluorescence from the spin-flipping $D$ transition \cite{Muller2014}. The device used in this work has $\Delta_{GS} = 554$ GHz.}
\label{SIfig:StrainT1Dep}
\end{figure}

Our results for the temperature dependence of $T_1^{-1}$ for several SiVs in the same device are shown  in Fig. \ref{SIfig:StrainT1Dep}. These SiVs have a range of ground state splittings $\Delta_{GS}$ arising from the inhomogeneous distribution of local strain within the same nanocavity. We normalize the $T_1$ for each SiV to their value at the base temperature of 100 mK $(T_{1,0})$ since the absolute value of $T_1$ for each SiV can differ due the how well aligned the magnetic field is to their symmetry axes, but we are only interested in the scaling of $T_1$ with temperature. The $T_1$ behavior is fitted with the single-phonon and two-phonon process scaling behaviors as described in \cite{Meesala2018}. 

As expected, the SiVs with larger ground state splitting, including the main SiV used in this work ($\Delta_{GS} = 554$ GHz), exhibit the smallest $T_1$ drop with increasing temperature in this temperature range. Furthermore, the temperature at which two-phonon processes begins to dominate -- as indicated by the kink in the slope that precedes a rapid decrease in $T_1$ -- increases with increasing $\Delta_{GS}$ as predicted from the theory. In particular, we find another device with  $\Delta_{GS} = 416$ GHz that is also highly strained and has excellent $T_1$ behavior up to 1.5 K, further confirming that the large $\Delta_{GS}$ from strain is a key determinant towards the high-temperature operation of our current device via the suppression of two-phonon processes.

\section{Distribution of SiV Strain Across Devices}

\begin{figure}[t]
\centering
\includegraphics[trim=0cm 13.5cm 9cm 0cm]{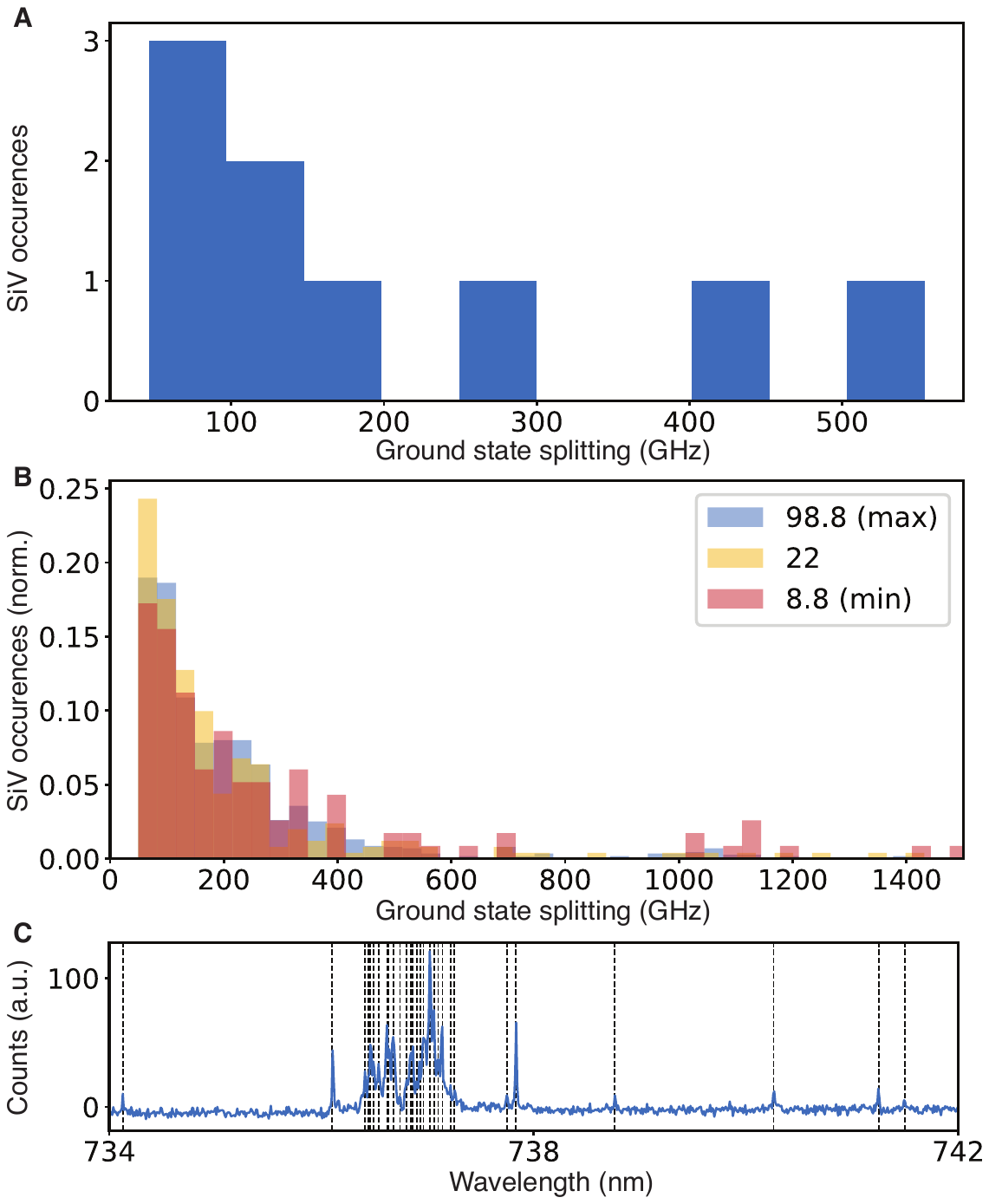}
\caption{(\textbf{A}) Distribution of $\Delta_{GS}$ measured in the same device as the main SiV used in this paper using resonant excitation. (\textbf{B}) Distribution of $\Delta_{GS}$ in a different device using off-resonant excitation. The different histograms correspond to the different numbers of average SiVs per cavity used during the fitting procedure. (\textbf{C}) Photoluminescence spectrum of one of the measured cavities with matched peaks marked with vertical lines.}
\label{SIfig:StrainDistr}
\end{figure}

We characterize the approximate ground state splitting distribution in our devices to estimate the feasibility of preselecting SiVs in nanocavities for high strain. First, we perform direct measurements of the ground state splitting with resonant excitation for several SiVs that were found in the same device as the main SiV used in this paper (Fig. \ref{SIfig:StrainDistr}A). We send light resonant with an individual SiV's C transition and measure the outgoing light from the cavity with a spectrometer at zero magnetic field. The measured spectrum shows two peaks corresponding to the SiV C and D transitions, and the difference in frequency between these peaks is the ground state splitting value for the given SiV. We measure 9 different SiVs in our device and find 2 highly strained SiVs with $\Delta_{GS} >$ 400 GHz.

For a more statistically significant sample of the strain distribution, we perform a large-scale spectroscopy of $\Delta_{GS}$ on a separate chip with 12 cavities that was fabricated in an identical manner. The measurement is taken in a 4 K closed-cycle helium cryostat (Montana Instruments) with no magnetic field applied. The SiVs are illuminated off-resonantly at 520 nm through a free-space port, after which fluorescence counts are collected confocally and sent to a spectrometer (Fig. \ref{SIfig:StrainDistr}C). The observed peaks are then grouped into sets of 4 peaks corresponding to the four A--D transitions of single SiVs fulfilling the requirements $\omega_A-\omega_B = \omega_C-\omega_D > 48$ GHz (minimum ground state splitting for an unstrained SiV) and $\omega_A-\omega_C = \omega_B-\omega_D > 259$ GHz (minimum excited state splitting for an unstrained SiV) \cite{HeppThesis}. 

The number of SiVs per cavity is not known with certainty, but for a given cavity we can find an estimated minimum number (so that all peaks that can possibly be matched with at least 1 SiV are assigned a match) and a maximum number of SiVs (all allowed sets of 4 peaks are assigned to a separate SiV). For a given choice of the mean number $\mu$ of SiVs per cavity, we can obtain a sample of the estimated $\Delta_{GS}$ distribution by first sampling a value $n$ for the number of SiVs per cavity from the normal distribution $n\sim\mathcal{N}(\mu, \sigma=2)$ (clipped within the bounds explained above), and then sampling a possible assignment configuration of $n$ SiVs to the peaks in the photoluminescence spectra, from which their $\Delta_{GS}$ may be extracted. This sampling procedure is repeated 10,000 times for each choice of the mean number of SiVs per cavity. 

Fig. \ref{SIfig:StrainDistr}B shows the resulting ground state splitting distributions for the minimum (8.8) and maximum (98.8) number of SiVs assigned per cavity during the fitting procedure. We also show the results for 22 SiVs per cavity, which is an estimate of the actual value based on resonant excitation measurements performed on the chip used in this experiment. We find that the proportion of SiVs with large ground state splitting ($\Delta_{GS} > 400$ GHz) is $12.2\pm1.4\%$ for 22 SiVs per cavity, and the lowest estimate of this proportion is $11.0\pm1.1\%$ when assigning the maximum number of SiVs per cavity.

\section{Spin-fluctuator model and motional averaging}

\begin{figure}[t]
\centering
\includegraphics[trim=0cm 22.5cm 13cm 0cm]{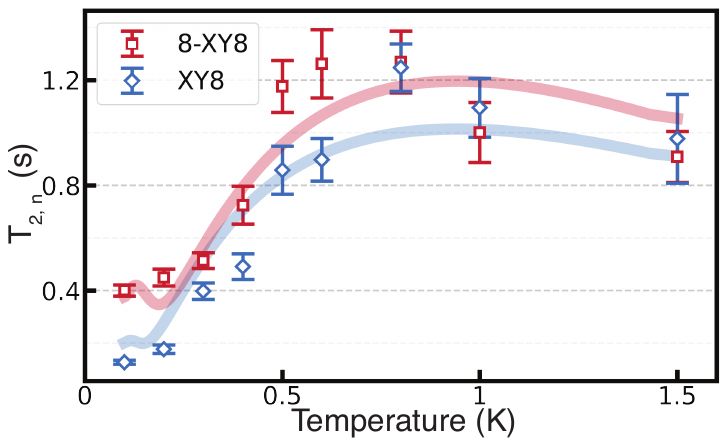}
\caption{Coherence time of the nuclear spin as a function of temperature. Solid line is a fit to the decoherence model of the spin environment.}
\label{SIfig:nuc-temp-t2}
\end{figure}

The temperature dependence of the coherence times for the electron and nucleus was investigated in the main text. Here, we present the models of surrounding spin and phonon environment that were used to fit the observed data. The electron coherence time is primarily determined by coupling to a non-Markovian magnetic noise bath and dephasing from phonon excitation processes. The phonon excitation processes we include are a resonant 1-phonon process and a resonant 2-phonon process with rate scalings taken from \cite{Meesala2018}. The depolarization timescale $T_{1,e}$ is much larger than any measured $T_{2,e}$, so we assume it does not significantly contribute to dephasing. The resulting temperature dependence of the electron coherence time that we used for the fitting is:
$$ T_{2,e}^{\mathrm{tot}}(T) = \left ( T_{1-\mathrm{phonon}}^{-1}(T) + T_{2-\mathrm{phonon}}^{-1}(T) + T_{\mathrm{noise bath}}^{-1} \right )^{-1}$$

On the other hand, for the dependence of the nuclear coherence time on temperature, we incorporate the contributions from three separate decoherence processes. First, depolarization of the electron causes dephasing of the nucleus due to the strong hyperfine interaction between the electron and the nucleus resulting in differential phase accumulation. The rate of this process can be measured through the electron $T_{1,e}$ (Fig. \ref{SIfig:StrainT1Dep}). Second, we expect the nuclear coherence to be affected by a bath of $^{13}$C nuclear spins. As the $^{13}$C nuclear spin couples very weakly to phonons \cite{ONeill2021}, this spin bath is modeled to be temperature-independent. Finally, we add another spin bath that is modelled as a bath of two-level spins with phonon-addressable transitions. We model this bath as causing discrete frequency fluctuations of the nuclear spin with a Monte Carlo simulation \cite{Sagi2010}, with the rate of these fluctuations scaling with the temperature-dependent phonon density at the characteristic transition frequencies of the bath \cite{Meesala2018}. Through the effects of motional averaging \cite{Anderson1954, Jiang2008}, this last bath leads to the observed increase in nuclear coherence with temperature. This is because as the temperature increases, so does the rate of fluctuations of the phonon-sensitive spin bath; when this fluctuation timescale becomes much smaller the decoupling sequence timescale, the fluctuations average out and cause the nucleus to feel a weaker averaged perturbation from the bath. This leads to the coherence time increase as seen in Fig. \ref{SIfig:nuc-temp-t2}. Combining all processes, the resulting temperature dependence of the nuclear coherence time is:
$$ T_{2,\mathrm{tot}}(T) = \left ( T_{1,e}^{-1}(T) + T_{2,\mathrm{^{13}C}}^{-1} + \sum_i T_{2,\mathrm{ph}}^{-1}(T, \omega_i) \right )^{-1}$$
where $T_{1,e}(T)$ is the electron coherence time, $T_{2,\mathrm{^{13}C}}$ is the temperature-independent dephasing from $^{13}$C nuclear spins, and  $T_{2,\mathrm{ph}}(T, \omega_i)$ is dephasing from a spin bath of two-level systems with a phonon-sensitive transition at frequency $\omega_i$. We fit the temperature coherence data in Fig. \ref{SIfig:nuc-temp-t2} and find two characteristic phonon-addressable transition frequencies at 26 GHz and 34 GHz, which could potentially correspond to reconfigurations of microstrain in the diamond lattice, or orbital state splittings of nearby defects. 

\section{Optimization of the phase readout}

Many quantum networking protocols, including entanglement distillation and swapping with nondeterministic heralded entanglement schemes, require multiple readouts of the electron state in the middle of the experimental sequence while preserving the state of the nucleus \cite{Kalb2018}. To maximize the number of readouts and their fidelity, we can optimize the cavity-QED parameters of the SiV in the nanophotonic cavity together with the readout laser parameters.

\begin{figure}[t]
\centering
\includegraphics[trim=0cm 17.5cm 7cm 0cm]{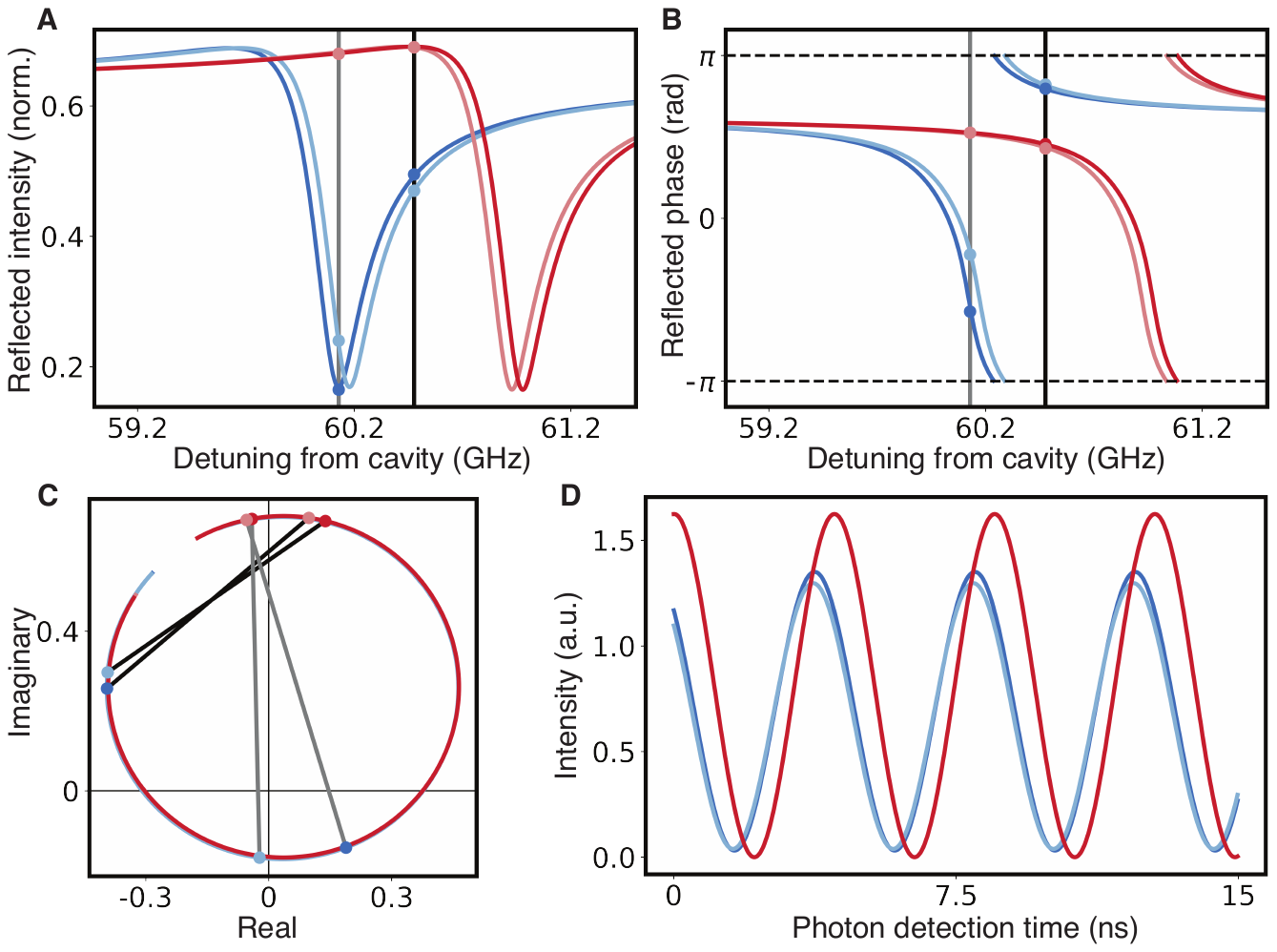}
\caption{Cavity reflection spectra for two-qubit states: $\ket{\downarrow_e \downarrow_n}$, $\ket{\downarrow_e \uparrow_n}$, $\ket{\uparrow_e \uparrow_n}$ and $\ket{\uparrow_e \downarrow_n}$ in dark/light blue and dark/light red. Cavity-QED parameters was chosen for a typical device with $\{g = 2\pi \times 4, \kappa_{in} = 2\pi \times 190, \kappa_{tot} = 2\pi \times 250, \Delta_a = 60,\Delta_e = 0.8,\Delta_n = 0.05, \gamma= 0.16 \}$ GHz. (\textbf{A, B}) Magnitude and phase of a reflected photon. Gray line: Frequency of maximum amplitude difference between electron states, used for resonant readout. Black line: Optimized frequency with maximum phase-space separation for the electron states and minimum nuclear decoherence, used for phase readout. (\textbf{C}) Complex reflection amplitude for the four different states as the laser frequency is varied, which forms a circle in the complex plane. Points indicate the resonant readout frequency (gray) and optimized nuclear state-preserving readout frequency (black). (\textbf{D}) Phase-measurement signal where the optimum readout frequency for the given cavity-QED parameters is interfered with another readout frequency far from SiV resonance.}
\label{SIfig:phase-amp-cqed}
\end{figure}

A useful optimization metric as a function of laser frequency $f$ is the ratio of the trace distances of electron states over the trace distances of the nuclear states since the trace distances are proportional to the maximum distinguishability between the respective spin states. We only consider the reflected photonic states for the electron since this is the only signal we can collect for state readout. On the other hand, we calculate the overlap over all three paths: reflection, scattering, and transmission for the nuclear trace distance. The key is to maximize the electron state information and minimize the nuclear state disturbance by each measurement \cite{fuchs1996distinguishability}.

The measurement constitutes sending in a coherent state to the cavity as an input and getting spin-state dependent output in three ports reflection, transmission and scattering:
\begin{align*}
    \text{Before interaction:} & \\
    &\ket{\psi_0} = \ket{\alpha} \otimes (\ket{\uparrow} + \ket{\downarrow}) \\
    \text{After interaction:} & \\
    &\ket{\psi_1} = \ket{R_\uparrow \alpha} \otimes \ket{\uparrow} + \ket{R_\downarrow \alpha} \otimes \ket{\downarrow} \\
    \label{eqn:cohReadout}
\end{align*}
where we can only collect the signal from the reflected port. The output coherent states are pure states that encode the information about the spins in our system. Since no two coherent states are orthogonal, the trace distance between the output states are not zero: $d_{{\uparrow \downarrow}_e} = \sqrt{1 - \braket{R_\downarrow \alpha}{R_\uparrow \alpha}^2 } \neq 0$. This results in an upper bound of maximum probability to distinguish two electron spin states using the reflected coherent sates:
\begin{align*}
    d_{{\uparrow \downarrow}_e} = \sqrt{1 - \braket{R_\downarrow \alpha}{R_\uparrow \alpha}_e^2 } 
\end{align*}
The overlap $\braket{R_{\downarrow} \alpha}{R_{\uparrow} \alpha}$ determines how well we can distinguish the spin states in our measurement. The farther apart the spin-dependent coherent states are in phase space, the more distinguishable the spin states are in the measurement. There are two ways to change the phase space distances: the first one is to pick a frequency at which the two cavity reflection coefficients are farthest from each other on the circle in phase space traced out by the complex reflection coefficients as the laser frequency is varied (Fig. \ref{SIfig:phase-amp-cqed}C), and the second one is to increase the amplitude $\abs{\alpha}$, which scales the diameter of the circle by shining more photons onto the cavity.

We need to include all output ports for the nuclear trace distance since decoherence is not dependent on our efficiency in collecting data from multiple ports.
\begin{align*}
    d_{\uparrow \downarrow_n} = \sqrt{1 - \braket{R_\downarrow \alpha}{R_\uparrow \alpha}_n^2 \braket{T_\downarrow \alpha}{T_\uparrow \alpha}_n^2 \braket{S_\downarrow\alpha}{S_\uparrow \alpha}_n^2 } 
\end{align*}
where $d_{{\uparrow \downarrow}_n}$ is the trace distance between the reflected photonic states for the two nuclear states. Finally, our optimization function is the ratio of the probabilities to distinguish the electron state over nuclear states as a function of the laser frequency $f$: 
\begin{align*}
    Optimizer(f) & = \frac{d_{\uparrow \downarrow_e}}{d_{\uparrow \downarrow_n}} \\
    & =\frac{\sqrt{1 - \braket{R_\downarrow \alpha}{R_\uparrow \alpha}^2_e }}{ \sqrt{1 - \braket{R_\downarrow \alpha}{R_\uparrow \alpha}_n^2 \braket{T_\downarrow \alpha}{T_\uparrow \alpha}_n^2 \braket{S_\downarrow \alpha}{S_\uparrow \alpha}_n^2 } }
\end{align*}
Note that the reflection coefficients for the electron spin-down and spin-up states also depend on the nuclear state, so the numerator implicitly includes an average over both nuclear states. The denominator similarly includes an implicit average over both electron states. 

For any given choice of cavity-QED and SiV parameters, we can minimize this optimizer to find an optimum choice of laser frequency for electron state readout while preserving nuclear coherence. We can then compute the number of photons required at this frequency to perform an electron state phase-based readout at 95\% fidelity, which gives us the nuclear decoherence per readout. This lets us compute the number of electron readouts possible before we experience a $1/e$ nuclear state decoherence, which we then optimize across all practical device and SiV parameters.

We find that systems where more electron readouts are possible for a given amount of nuclear decoherence typically: (I) have a larger $g$ and have over-coupled cavities. These parameters maximize the amount of incoming light directed into the reflected port, thus increasing the amount of information gained per photon sent; (II) have a large electron splitting $\Delta_e$ while keeping the nuclear state splitting $\Delta_n$ constant, which an be achieved by ramping up the magnetic field. However, operation at the correspondingly higher microwave frequencies have limiting factors, including signal delivery inefficiencies and an eventual reduction of $T_{1,e}$ due to an increased phononic density of states. In particular, for achievable parameters of $\{g = 2\pi \times 7, \kappa_{in} = 2\pi \times 180, \kappa_{tot} = 2\pi \times 250, \Delta_a$ (cavity detuning) $ = 100,\Delta_e = 1.5,\Delta_n = 0.05, \gamma= 0.16\}$ GHz, we expect to be able to do approximately 90 electron readouts at 95\% fidelity without postselection before we experience $1/e$ nuclear decoherence.

The optimum frequency above gives the best probe point if one is able to detect both amplitude and phase differences in the reflected signal. Practically, it may be more convenient to sacrifice some electron state distinguishability or nuclear state fidelity by purely optimizing for the amplitude contrast or phase difference between the reflected probe coherent states. This is especially useful for designing ``one-shot'' state readouts similar to our resonant readout or a homodyne readout.

\end{document}